\documentclass[aps,twocolumn]{revtex4-1}
\usepackage{graphicx}
\usepackage{amsmath}
\usepackage{amssymb}
\usepackage{bm}
\usepackage{color}

\begin{document}
\title{Machine-Learning Studies on Spin Models} 
\author{Kenta Shiina$^{1,2}$}
\email{16879316kenta@gmail.com}
\author{Hiroyuki Mori$^{1}$}
\author{Yutaka Okabe$^{1}$}
\email{okabe@phys.se.tmu.ac.jp}
\author{Hwee Kuan Lee$^{2,3,4,5}$}
\email{leehk@bii.a-star.edu.sg}
\affiliation{
$^1$Department of Physics, Tokyo Metropolitan University, 
Hachioji, Tokyo 192-0397, Japan \\
$^2$Bioinformatics Institute, 
Agency for Science, Technology and Research (A*STAR),
30 Biopolis Street, \#07-01 Matrix, 
138671, Singapore \\
$^3$School of Computing, National University of Singapore, 
13 Computing Drive, 
117417, Singapore \\
$^4$Singapore Eye Research Institute (SERI), 
11 Third Hospital Ave, 
168751, Singapore \\
$^5$Image and Pervasive Access Laboratory (IPAL), 
1 Fusionopolis Way, \#21-01 Connexis (South Tower), 
138632, Singapore
}

\def\l{\langle}
\def\r{\rangle}

\date{\today}

\begin{abstract}
With the recent developments in machine learning, 
Carrasquilla and Melko have proposed 
a paradigm that is complementary to the conventional approach 
for the study of spin models.  As an alternative to investigating 
the thermal average of macroscopic physical quantities, 
they have used the spin configurations for the classification 
of the disordered and ordered phases of a phase transition 
through machine learning. 
We extend and generalize this method. We focus on 
the configuration of the long-range correlation function 
instead of the spin configuration itself, which enables us to provide 
the same treatment to multi-component systems and the systems 
with a vector order parameter. 
We analyze the Berezinskii-Kosterlitz-Thouless (BKT) transition 
with the same technique to classify three phases: the disordered, 
the BKT, and the ordered phases. 
We also present the classification of a model 
using the training data of a different model. 
\end{abstract}

\maketitle


Numerical simulations, such as Monte Carlo methods, have been 
successfully employed in the study of phase transitions and 
critical phenomena~\cite{Landau}. 
In spin systems, the spin configurations are sampled using a stochastic 
importance technique, and the estimators for physical quantities, 
such as the order parameter and the specific heat, are evaluated 
for these samples. 

Several spin models have recently been studied through 
machine learning~\cite{Carrasquilla,Beach,Suchsland,Zhang,Rodriguez-Nieva}. 
Carrasquilla and Melko~\cite{Carrasquilla} proposed 
a paradigm that is complementary to the above approach. 
By using large data sets of spin configurations, 
they classified and identified a high-temperature paramagnetic phase 
and a low-temperature ferromagnetic phase. 
It was similar to image classification using 
machine learning. 
They demonstrated the use of fully connected and 
convolutional neural networks for the study 
of the two-dimensional (2D) Ising model and 
an Ising lattice gauge theory. 

In this study, we extend and generalize the method proposed by 
Carrasquilla and Melko~\cite{Carrasquilla}. 
First, instead of considering the spin configuration itself, 
we analyze the long-range correlation configuration, 
which will be explained later. From this analysis, 
we can evaluate the multi-component systems, such as the Potts model, 
and the systems with a vector order parameter, such as the XY model. 
We can identify identical configurations with the permutational symmetry 
or the rotational symmetry, which results in an efficient classification 
of phases.  Moreover, the inclusion of long-range correlation 
is helpful in the study of phase transition. 
Second, we investigate the Berezinskii-Kosterlitz-Thouless 
(BKT) phase \cite{Berezinskii1,Berezinskii2,kosterlitz,kosterlitz2}, 
described by a fixed line instead of a fixed point 
from the perspective of the renormalization group, 
using the same treatment as the paramagnetic-ferromagnetic phase 
transition.  By studying the 2D clock model, which is a discrete version 
of the XY model, we classify the paramagnetic-BKT-ferromagnetic transitions 
through machine learning. 

\section*{Model}

We enlist the models we analyze below. 
We consider a 2D Ising model on the square lattice, 
whose Hamiltonian is given as
\begin{equation}
 H = -J \sum_{\l ij \r} s_i s_j, \quad s_i = \pm 1.
\label{Ising}
\end{equation}
The summation is realized over the nearest-neighbor pairs, 
and periodic boundary conditions are imposed in numerical 
simulations. 

The Hamiltonian of the $q$-state Potts model \cite{Potts,Wu} 
is given by
\begin{equation}
 H = -J \sum_{\l ij \r} \delta_{s_i s_j}, \quad s_i = 1, 2, \cdots, q,
\label{Potts}
\end{equation}
where $\delta_{a b}$ is the Kronecker delta. 
The 2D ferromagnetic Potts model is known to exhibit a second-order 
phase transition for $q \le 4$ and a first-order phase transition 
for $q \ge 5$. The Potts model for $q=2$ is identical to the Ising model. 

The 2D spin systems with a continuous XY symmetry exhibit a unique 
phase transition called the BKT transition 
\cite{Berezinskii1,Berezinskii2,kosterlitz,kosterlitz2}. 
A BKT phase of a quasi long-range order (QLRO) exists, 
wherein the correlation function decays as a power law. 
Here, we consider the $q$-state clock model, which is a discrete version 
of the classical XY model. 
Its Hamiltonian is given by
\begin{equation}
   H = -J \sum_{\l ij \r} \cos (\theta_i - \theta_j), \quad
   \theta_i = 2\pi i/q, \ i = 1, 2, \cdots, q.
\label{clock}
\end{equation}
The 2D $q$-state clock model experiences a BKT transition 
for $q \ge 5$, whereas the clock model for $q=4$ comprises 
two sets of the Ising model and the 3-state clock model 
is equivalent to the 3-state Potts model. 
The clock model for $q=2$ is simply the Ising model.

We measure temperature in units of $J$.

\section*{Correlation configuration}

The correlation function in the Ising model, with a distance $r$, 
is given by
\begin{equation}
    g_i(r) = s_{i}s_{i+r}.
\end{equation}
It clearly assumes a value of $+1$ or $-1$.

In the case of the $q$-state Potts model, the correlation 
function is defined by
\begin{equation}
    g_i(r) = \frac{q \delta_{s_i s_{i+r}} - 1}{q-1}.
\end{equation}
It assumes a value of $+1$ or $-1/(q-1)$.

The correlation function $g_i(r)$ of the $q$-state clock model is 
\begin{equation}
    g_i(r) = \cos (\theta_i - \theta_{i+r}).
\end{equation}
It assumes a value between $+1$ and $-1$.

There are several types of symmetries in spin systems. 
A few different spin configurations are 
essentially identical, whereas they 
have the same correlation configuration.  

For phase transitions, it is preferable 
to include long-range correlations, which play 
an essential role in phase transitions. 
Because the longest distance in the finite-size systems 
of size $L$ with periodic boundary conditions is $L/2$, 
we consider the average value of the $x$-direction 
and the $y$-direction, that is, 
\begin{equation}
    g_i(L/2) = (s_{x,y}s_{x+L/2,y}+s_{x,y}s_{x,y+L/2})/2,
\label{g_i}
\end{equation}
for the Ising model. The same definitions are 
employed for other models.

We note that this type of correlation function was used 
along with the generalized scheme for the probability-changing 
cluster algorithm~\cite{tomita2002b}.


Using the Swendsen-Wang multi-cluster flip algorithm \cite{sw87} 
for updating spins, we generated the spin configurations for a given 
temperature $T$. 

The examples of the spin configurations $\{ s_i \}$ and correlation 
configurations $\{ g_i(L/2) \}$ for several models are shown 
in the supplementary information section. 
The plots of the 2D Ising model, the 2D 5-state Potts model, 
and the 2D 6-state clock model are shown in Fig.~S1, 
Fig.~S2, and Fig.~S3, 
respectively.

\section*{Machine-Learning Study}

\begin{figure}
\begin{center}
\includegraphics[width=8.6cm]{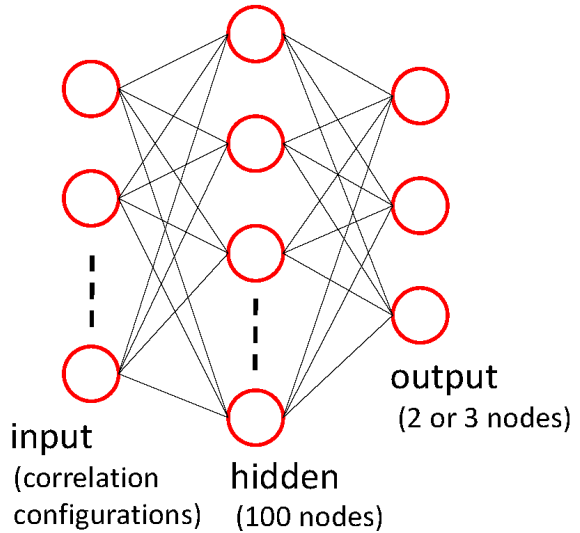}
\caption{
A schematic diagram of the fully connected neural network 
in the present simulation.
}
\label{fig:network}
\end{center}
\end{figure}

We have considered a fully connected neural network implemented 
with a standard TensorFlow library \cite{TF} using the 100-hidden 
unit model to classify the ordered and the disordered phases. 
For the input layer, we use correlation configurations $\{ g_i(L/2) \}$. 
A schematic diagram of the fully connected neural network 
in the present simulation is shown in Fig.~\ref{fig:network}. 
We have used a cross-entropy cost function supplemented with 
an $L2$ regularization term.  
The neural networks were trained using the Adam method \cite{Adam}. 

Typically, around 40,000 training data sets are used, and 
30,000 test data sets are used. Ten independent calculations were 
performed to provide error analysis. 
Although the exact transition 
temperatures $T_c$ are known for most of the models in the present study, 
we have not used the samples close to $T_c$ for the training data. 
We have assumed that the exact $T_c$ is not known. 

\begin{figure}
\begin{center}
{\bf a} \hspace*{6cm}

\includegraphics[width=8.6cm]{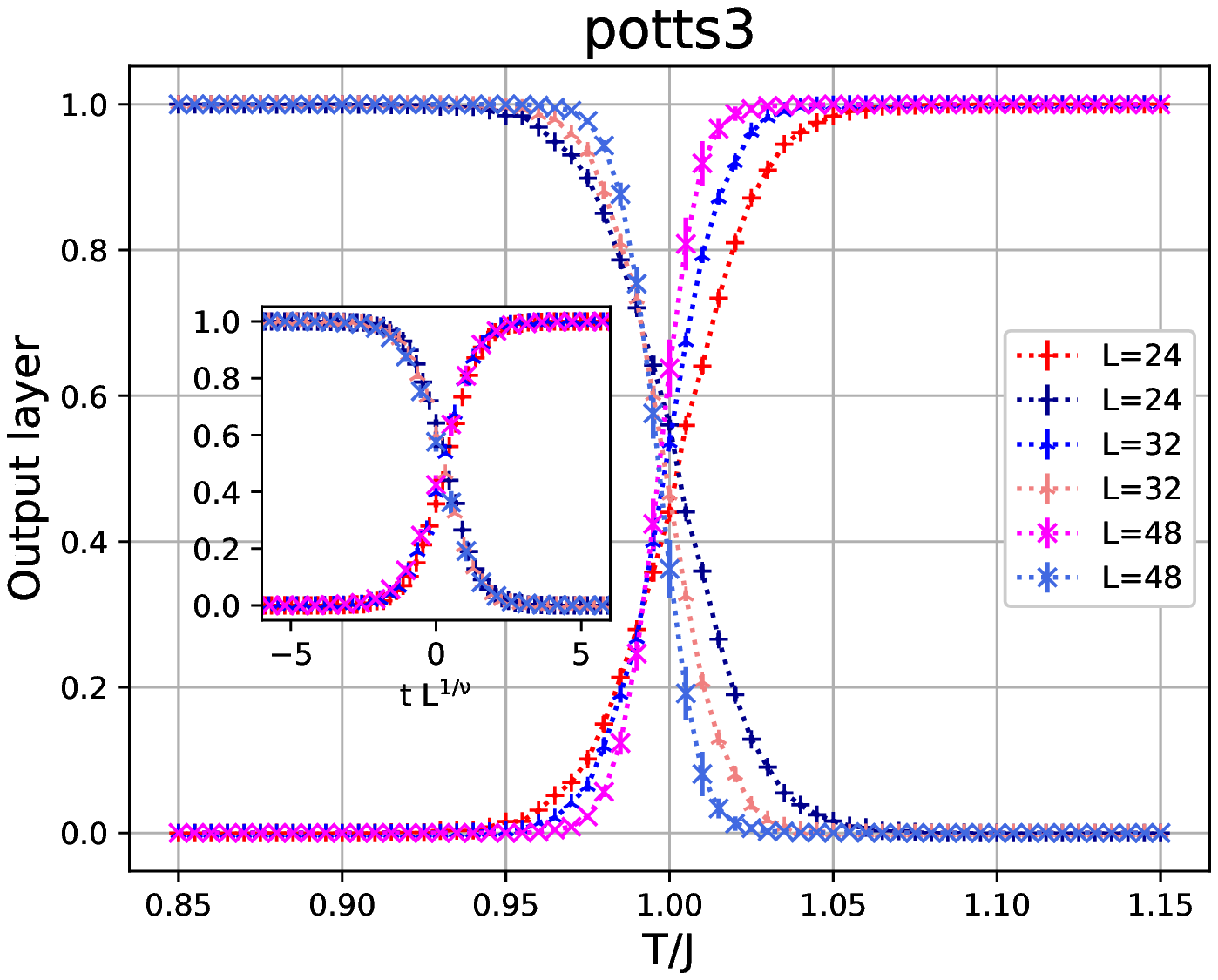}
\vspace*{2mm}

{\bf b} \hspace*{6cm}

\includegraphics[width=8.6cm]{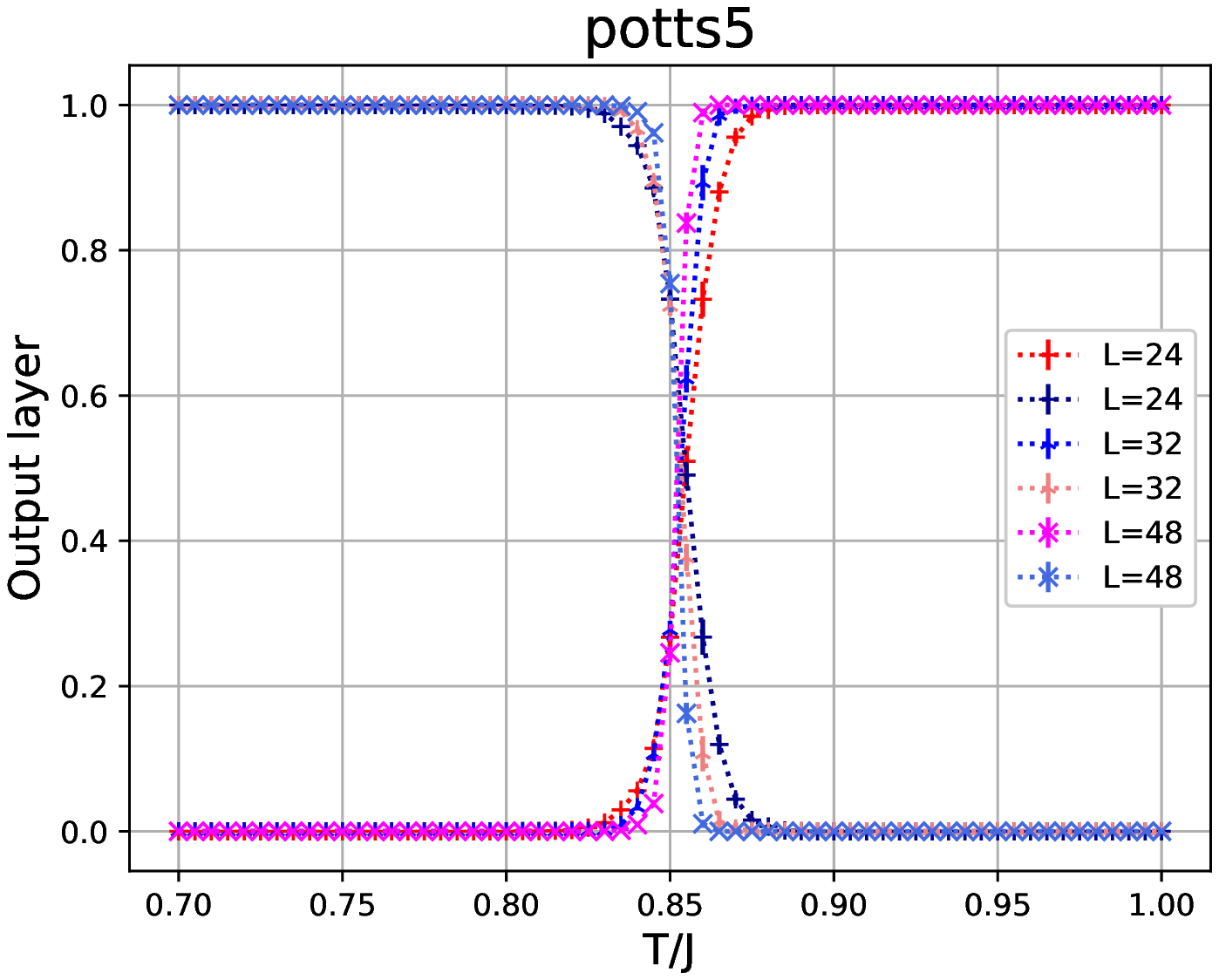}
\caption{
a) The output layer averaged over a test set as a function of $T$ 
for the 2D 3-state Potts model. The system sizes are $L$ = 
24, 32, and 48. 
The samples of $T$ within the ranges $0.85 \le T \le 0.94$ and 
$1.06 \le T \le 1.15$ are used for the training data. 
In the inset, the finite-size scaling plot is given, 
where the horizontal axis is chosen as $t L^{1/\nu}$ 
with $t=(T-T_c)/J$. The values of $T_c$ and $\nu$ are 
$T_c = 1/\ln(1+\sqrt{3}) = 0.995$ and $\nu$=5/6, respectively. 
b) The same plot for the 2D 5-state Potts model. 
The system sizes are the same. 
The samples of $T$ within the ranges $0.7 \le T \le 0.79$ and 
$0.91 \le T \le 1.0$ are used for the training data. 
}
\label{fig:output_potts3}
\end{center}
\end{figure}

We first analyzed the 2D 3-state Potts model. 
The output layer averaged over a test set as a function of $T$ 
for the 2D 3-state Potts model is shown in Fig.~\ref{fig:output_potts3}a. 
The probabilities of predicting the phases, the disordered or the ordered, 
are plotted for each temperature. 
The system sizes are $L$ = 24, 32, and 48. 
The samples of $T$ within the ranges $0.85 \le T \le 0.94$ and 
$1.06 \le T \le 1.15$ were used for the training data. 
The exact second-order transition temperature $T_c$ for this model 
is known as $1/\ln(1+\sqrt{3}) = 0.995$. 
We observed that the neural network could successfully classify 
the disordered and ordered phases. 
We give the finite-size scaling plot of the second-order 
transition \cite{fisher70} in the inset, 
where the horizontal axis is chosen as $t L^{1/\nu}$ with $t=(T-T_c)/J$ 
and the correlation-length exponent $\nu$.  
For the values of $T_c$ and $\nu$, we used the exact values, 
$T_c = 1/\ln(1+\sqrt{3}) = 0.995$ and $\nu$=5/6. 
We obtained very good finite-size scaling. 

We have presented the output layer averaged over a test set 
as a function of $T$ 
for the 2D 5-state Potts model in Fig.~\ref{fig:output_potts3}b. 
The system sizes are $L=$ 24, 32, and 48. 
The samples of $T$ within the ranges $0.7 \le T \le 0.79$ and 
$0.91 \le T \le 1.0$ were used for the training data. 
This model is known to exhibit the first-order transition at 
$T_c=1/\ln(1+\sqrt{5}) = 0.852$.  The transition is sharp 
compared with the Potts model for $q=3$ for the second-order 
transition. 

\begin{figure}
\begin{center}
{\bf a} \hspace*{6cm}

\includegraphics[width=8.6cm]{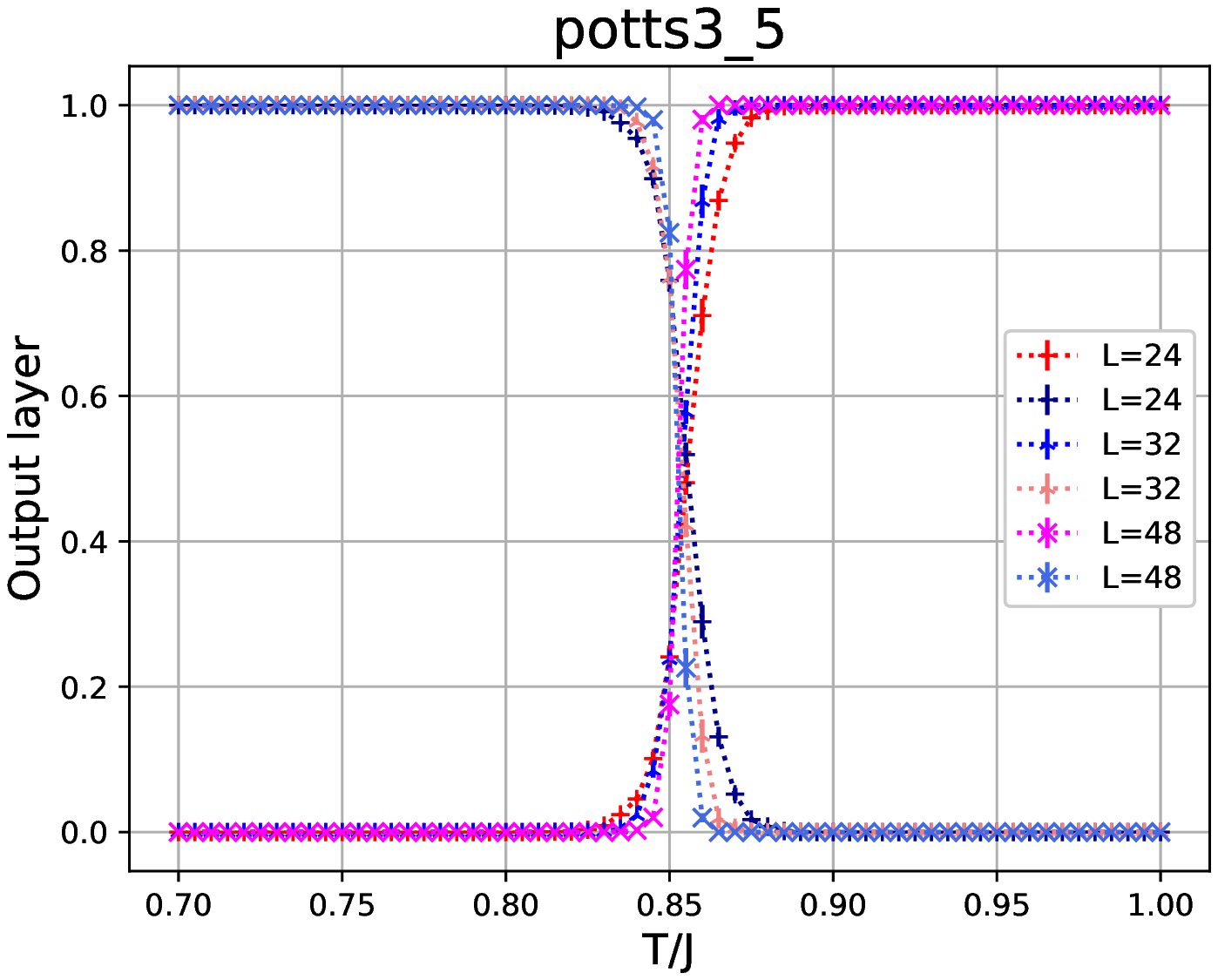}
\vspace*{2mm}

{\bf b} \hspace*{6cm}

\includegraphics[width=8.6cm]{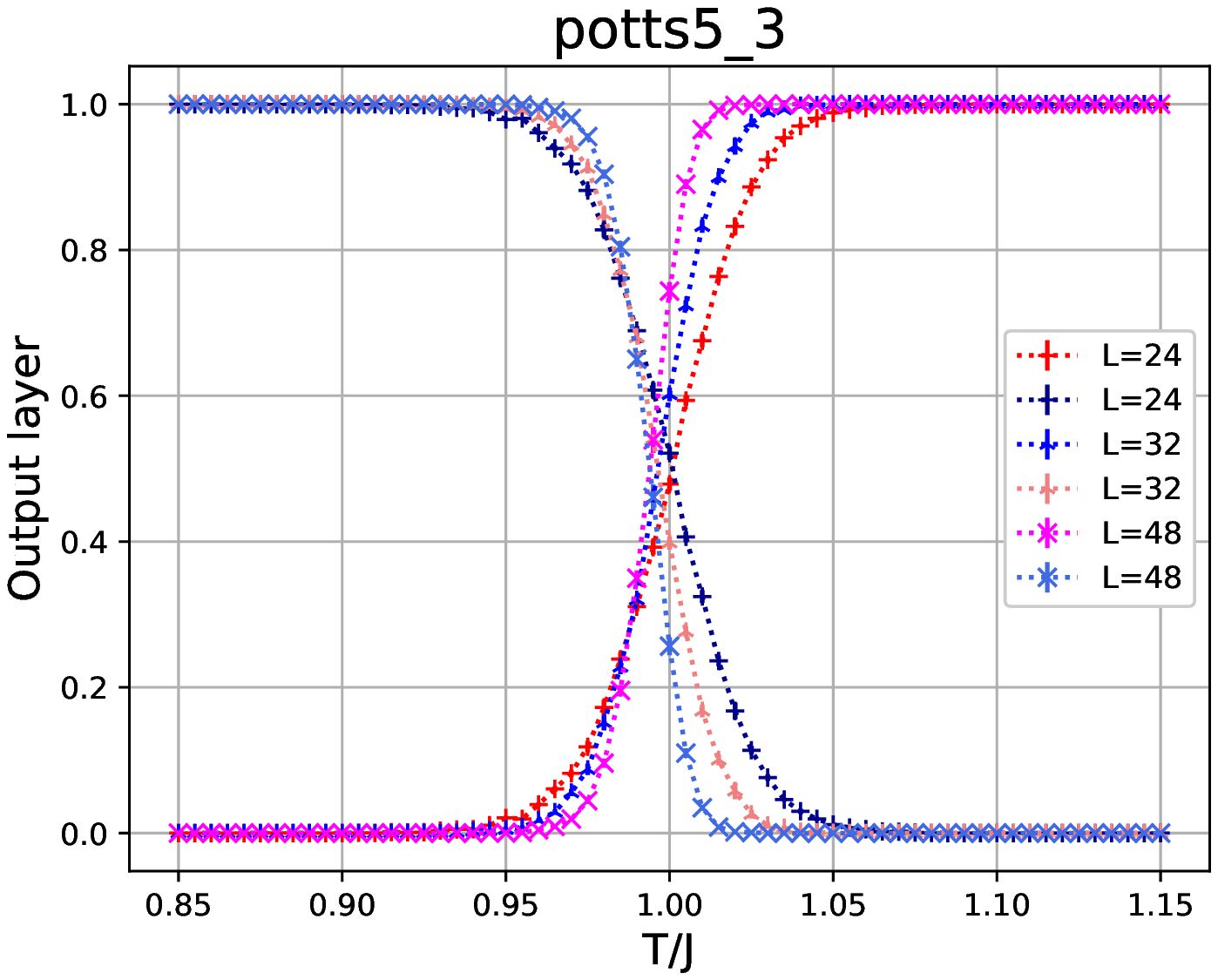}
\caption{
a) The output layer for the 5-state Potts model using the training data 
of the 3-state Potts model. 
b) The output layer for the 3-state Potts model using the training data 
of the 5-state Potts model.
}
\label{fig:output_potts3_5}
\end{center}
\end{figure}

It is instructive to use the training data obtained from the 
3-state Potts model for the classification of the phases 
of the 5-state Potts model. 
The output layer for the 5-state Potts model using the training data 
of the 3-state Potts model is given in Fig.~\ref{fig:output_potts3_5}a.  
It successfully reproduces the sharp transition  of the 5-state Potts model 
at $T_c=1/\ln(1+\sqrt{5}) = 0.852$. 
The plot of the opposite direction, that is, the output layer 
obtained for the 3-state Potts model using the training data of 
the 5-state Potts model is given in Fig.~\ref{fig:output_potts3_5}b. 
It reproduces the transition of the 3-state Potts model 
at $T_c=1/\ln(1+\sqrt{3}) = 0.995$. The order of the transition 
for the 3-state Potts model is second order, 
whereas that for the 5-state Potts model is first order.  
However, the training data of one model successfully 
reproduces the classification of the other model. 

\begin{figure}
\begin{center}
{\bf a} \hspace*{6cm}

\includegraphics[width=8.6cm]{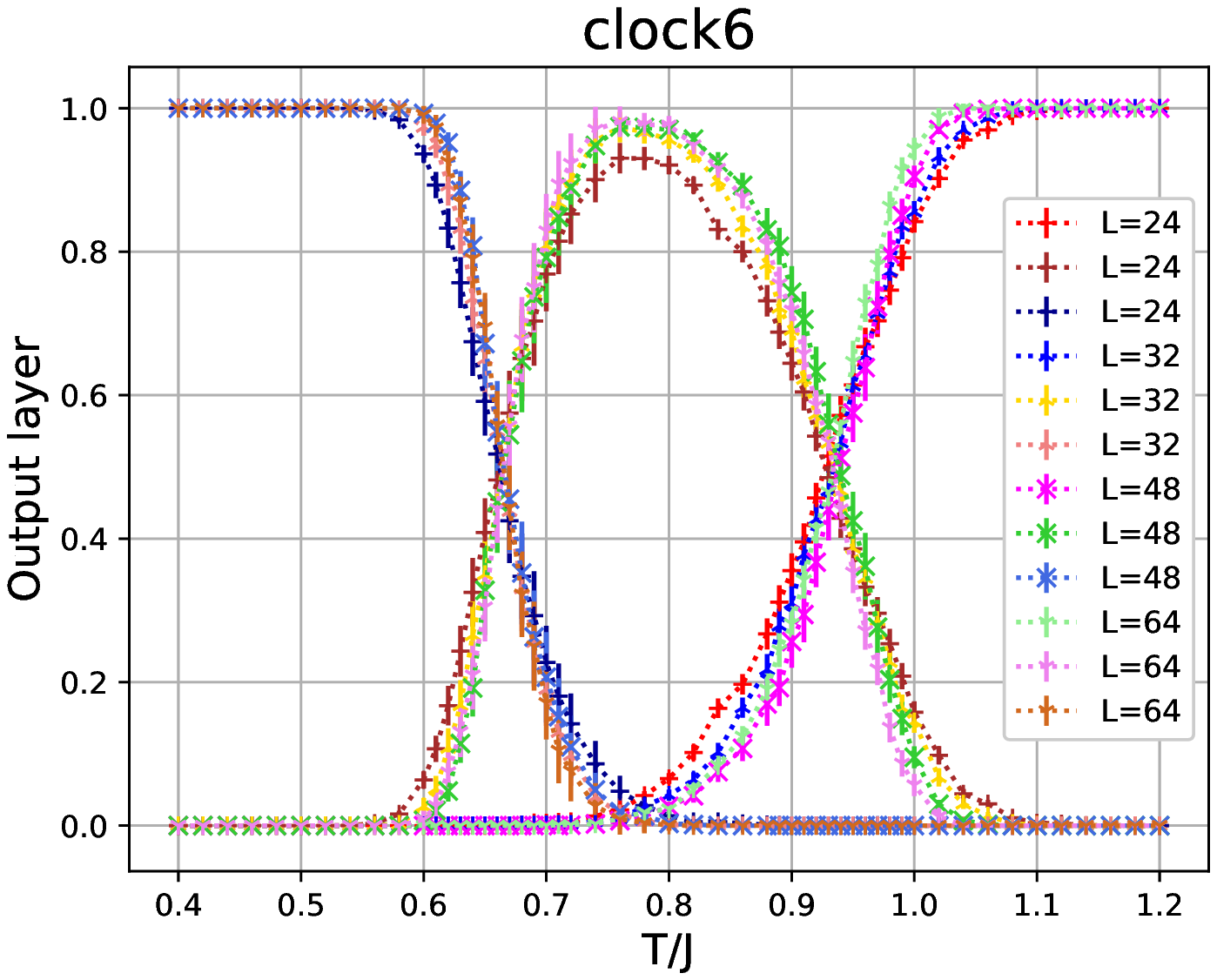}
\vspace*{2mm}

{\bf b} \hspace*{6cm}

\includegraphics[width=8.6cm]{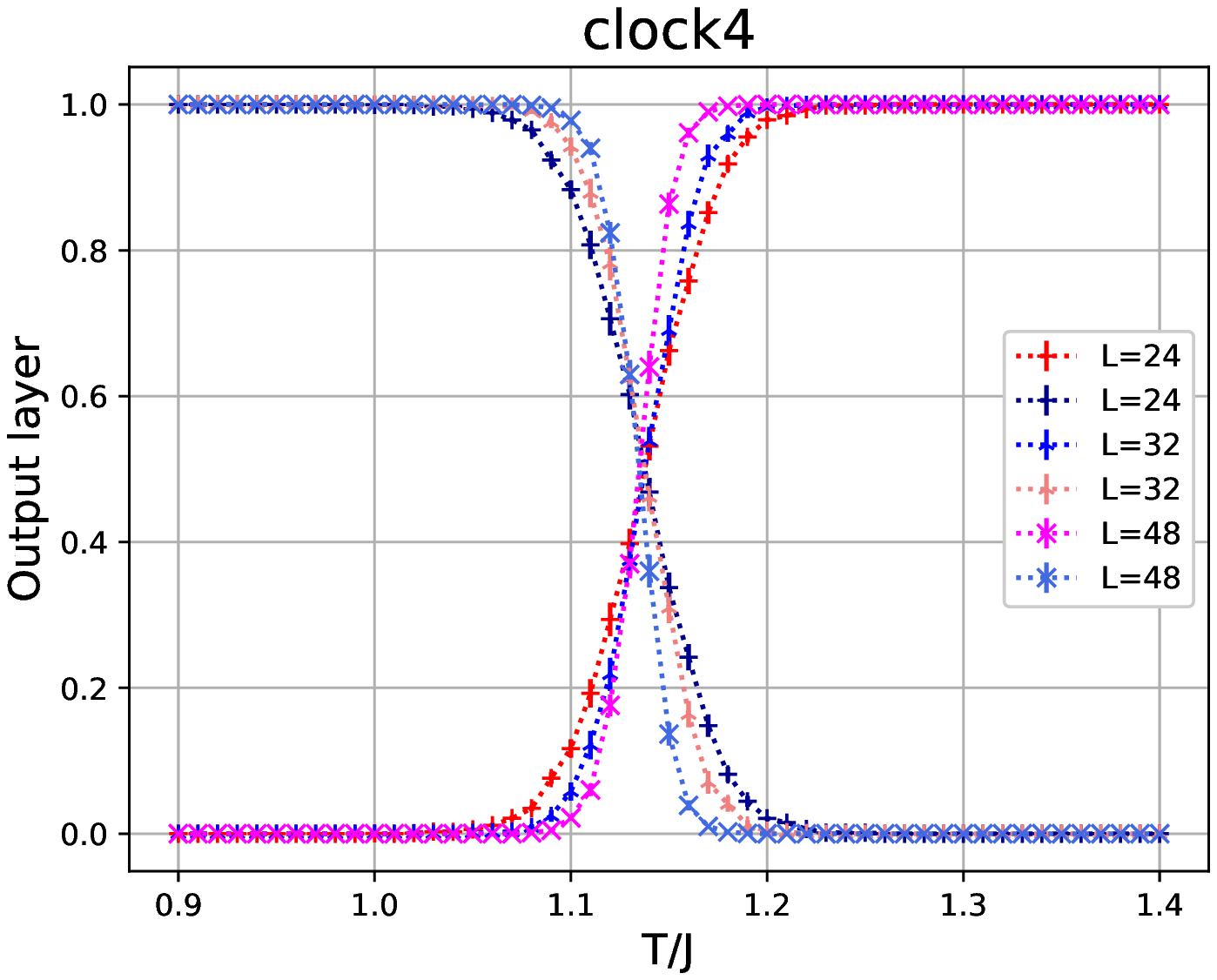}
\caption{
a) The output layer averaged over a test set as a function of $T$ 
for the 2D 6-state clock model. The system sizes are $L$ = 
24, 32, 48, and 64. 
The samples of $T$ within the ranges $0.4 \le T \le 0.64$, 
$0.77 \le T \le 0.83$, and 
$0.96 \le T \le 1.2$ are used for the training data. 
b) The same plot for the 2D 4-state clock model. 
The samples of $T$ within the ranges $0.9 \le T \le 1.06$ and 
$1.2 \le T \le 1.4$ are used for the training data. 
}
\label{fig:output_clock6}
\end{center}
\end{figure}

We have considered the 2D $q$-state clock model next. 
Because of the discreteness, there are two transitions for $q \ge 5$. 
One is a higher BKT transition, $T_2$, between the disordered phase 
and the BKT phase of QLRO, and the other is a lower transition, $T_1$, 
between the BKT phase and the ordered phase. 
The recent numerical estimates of $T_1$ and $T_2$ for the 
6-state clock model are 0.701(5) and 0.898(5), respectively \cite{Surungan}. 
The output layer averaged over a test set as a function of $T$ 
for the 2D 6-state clock model is shown in Fig.~\ref{fig:output_clock6}a. 
The system sizes are $L$ = 24, 32, 48, and 64. 
The samples of $T$ within the ranges $0.4 \le T \le 0.64$, 
$0.77 \le T \le 0.83$, and 
$0.96 \le T \le 1.2$ were used for the low-temperature, 
mid-range temperature, and high-temperature training data, respectively. 
Figure~\ref{fig:output_clock6}a shows the classification 
into the three phases. 
We estimate the size-dependent $T_{1,2}(L)$ from the point 
that the probabilities of predicting two phases are 50\%. 
The estimates of $T_1(L)$ and $T_2(L)$, in the range of $24 \le L \le 64$, 
are around 0.66-0.67 and 0.93-0.94, respectively. 
The correlation length at the BKT transitions diverges rapidly, 
as given below, 
\begin{equation}
   \xi \propto \exp (c/\sqrt{|t|})
\end{equation}
with $t = (T - T_{1,2})/T_{1,2}$, which is  both below $T_1$ and above $T_2$. 
Finite-size effects result in a wider prediction 
of the BKT phase for smaller sizes.  
Size effects become smaller gradually with $\ln L$. 
In the conventional Monte Carlo study of the BKT transition, 
the helicity modulus was calculated, and the size-dependent $T_2(L)$ 
can be estimated from the intersection with the straight line, 
$(2/\pi)*T$, the universal jump \cite{Weber,Harada}. 
The numerical estimates of $T_2(L)$ are 0.935 ($L=24$), 
0.929 ($L=32$), 0.925 ($L=48$), and 0.921 ($L=64$), 
which slowly converge to 0.898 in the infinite $L$ limit \cite{Surungan}. 
The present estimates of finite-size $T_2$ are compatible 
with the universal jump analysis, although the systematic 
size dependence is hided because of statitical errors. 
The situation for $T_1$ is the same. 
Thus, Fig.~\ref{fig:output_clock6}a clearly shows 
the behavior of the three phases. 

It is interesting to investigate the relation between 
the BKT transition and the second-order transition. 
For this purpose, we have examined the 4-state clock model. 
This model is equivalent to two sets of the Ising model; 
it has a single second-order transition at 
$T_c = 1/\ln(1+\sqrt{2}) = 1.135$. 
The output layer averaged over a test set as a function of $T$ 
for the 2D 4-state clock model is given in Fig.~\ref{fig:output_clock6}b. 
The samples of $T$ within the ranges $0.9 \le T \le 1.06$ and 
$1.2 \le T \le 1.4$ were used for the training data. 

\begin{figure}
\begin{center}
\includegraphics[width=8.6cm]{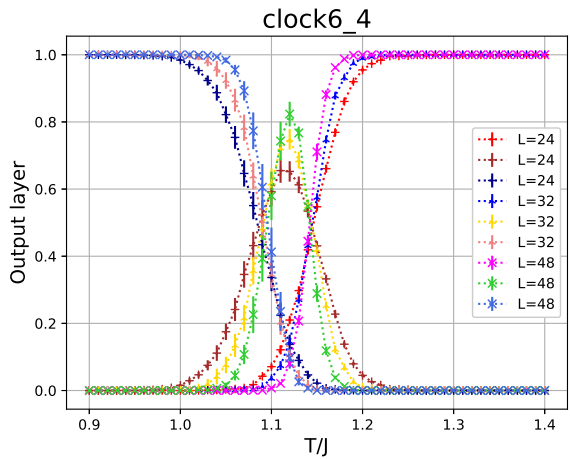}
\caption{
The output layer for the 4-state clock model using the training data 
of the 6-state clock model.
}
\label{fig:output_clock6_4}
\end{center}
\end{figure}

We investigate the result of using the training data of the 6-state clock 
model for the classification of the 4-state clock model. 
We present the output layer for the 4-state clock model 
as a function of $T$ using the training data of 
the 6-state clock model in Fig.~\ref{fig:output_clock6_4}. 
The phases of the 4-state clock model are classified into 
the ordered and disordered phases with the expected $T_c$ around 1.135. 
However, the narrow region near $T_c$ is regarded as the BKT phase. 
It is an indication that the BKT phase with a fixed line is the same as 
the critical phase of the second-order transition with a fixed point. 
The figure indicates that the critical region becomes narrower 
as the system size increases. 


\section*{Summary and discussion}

We reported a machine-learning study on several spin models 
to study phase transitions.  
We considered the configuration of a long-range 
spatial correlation instead of the spin configuration itself. 
By doing so, we provided a similar treatment to various spin models 
including the multi-component systems and the systems 
with a vector order parameter. 
We successfully classified the disordered and the ordered phases, 
along with the BKT type topological phase. 
We showed a good finite-size scaling plot for the second-order 
transition. 

Using the training data of the second-order transition system of 
the 3-state Potts model, we reproduced the phase classification 
of the first-order transition of the 5-state Potts model. 
The phase classification of the opposite direction was also successful. 
We achieved the phase classification of the second-order 
transition of the 3-state Potts model using the training data 
of the 5-state Potts model. 
Using the training data of the BKT transition system for 
the 6-state clock model, we elucidated the role of the critical phase 
of the second-order transition of the 4-state clock model. 
It is a direct demonstration that explains that the phase with a fixed line, 
whose spatial decay is an algebraic one, has the same structure 
as the critical phase of the second-order 
transition with a fixed point. 

The present treatment of machine-learning study is generalized, and 
can be applied to various systems including quantum spin systems. 
It will be interesting to study the universal behavior of 
the topological phase of BKT type. 
There are sometimes implicit symmetries in the models of physics. 
Universality appears in totally different systems. 
The 3-state antiferromagnetic square lattice Potts model 
with a ferromagnetic next-nearest-neighbor interaction 
is an example. This model was studied by Otsuka {\it et al.} \cite{Otsuka} 
using the level-spectroscopy method, where they presented two BKT transitions 
and the universality of the 6-state ferromagnetic clock model. 
The machine-learning study on this model is in progress, 
and it is expected to be reported in the future. 


\vspace*{1cm}
\section*{Acknowledgments}

This work was supported by a Grant-in-Aid for Scientific Research 
from the Japan Society for the Promotion of Science, Grant Number JP16K05480,
Tokyo Metropolitan University, Japan, and
the Biomedical Research Council of A*STAR
(Agency for Science, Technology and Research), Singapore.
KS is grateful to the A*STAR Research Attachment Programme (ARAP) of Singapore 
for financial support. 

\end{document}


\title{Machine-Learning Studies on Spin Models: 
Supplementary information} 
\author{Kenta Shiina}
\author{Hiroyuki Mori}
\author{Yutaka Okabe}
\author{Hwee Kuan Lee}

\def\l{\langle}
\def\r{\rangle}


\maketitle

\section*{Examples of the spin configurations 
and correlation configurations}

\setcounter{figure}{0}
\renewcommand{\figurename}{FIG.~S}

\begin{figure*}
\begin{center}

{\bf a} \hspace*{5.6cm}{\bf b}\hspace*{5.6cm}{\bf c}\hspace*{3.6cm}

\vspace*{1mm}

\includegraphics[width=5.6cm]{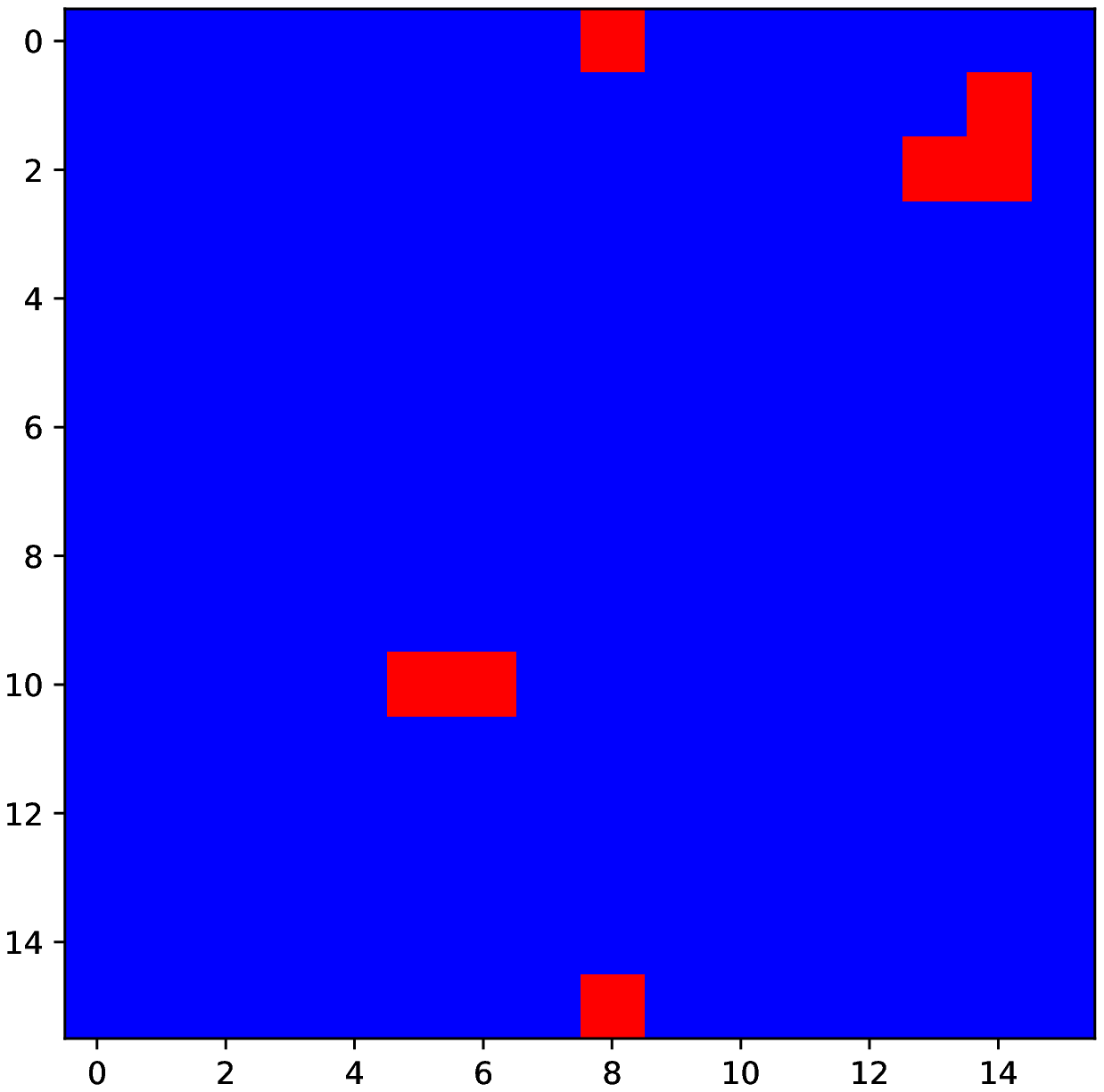}
\includegraphics[width=5.6cm]{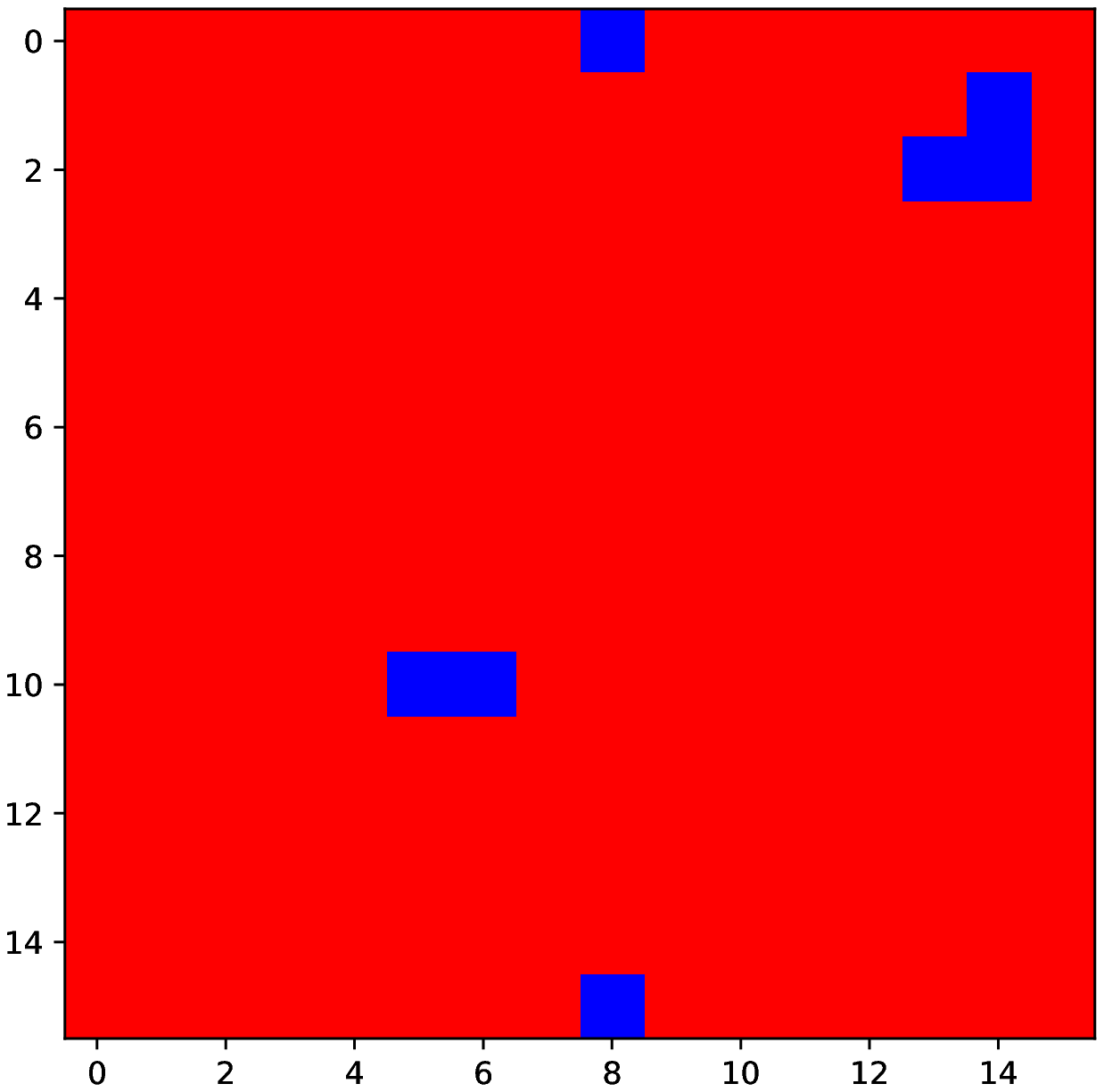}
\includegraphics[width=5.6cm]{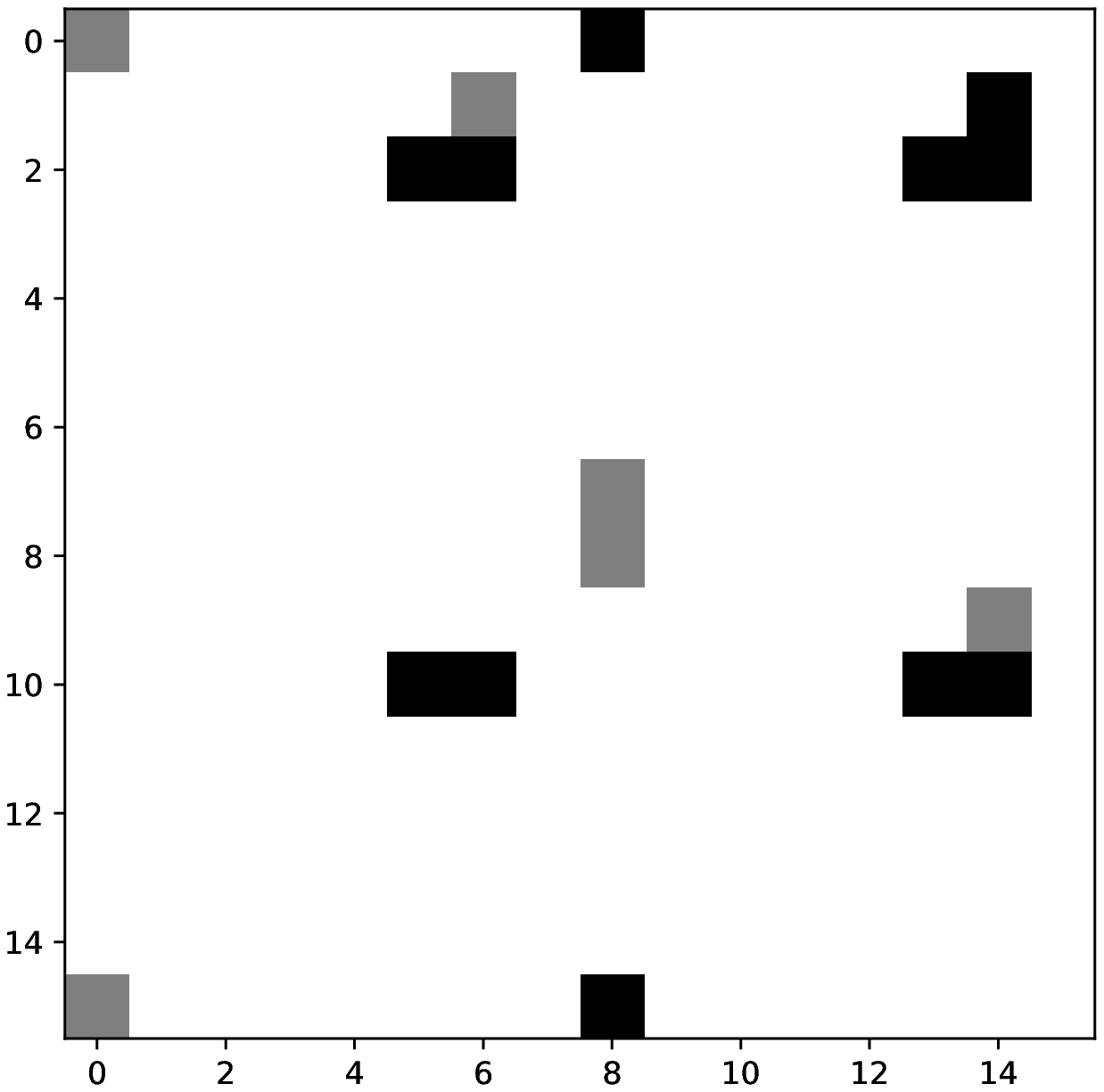}

\vspace*{4mm}

{\bf d} \hspace*{5.6cm}{\bf e}\hspace*{5.6cm}{\bf f}\hspace*{3.6cm}

\vspace*{1mm}

\includegraphics[width=5.6cm]{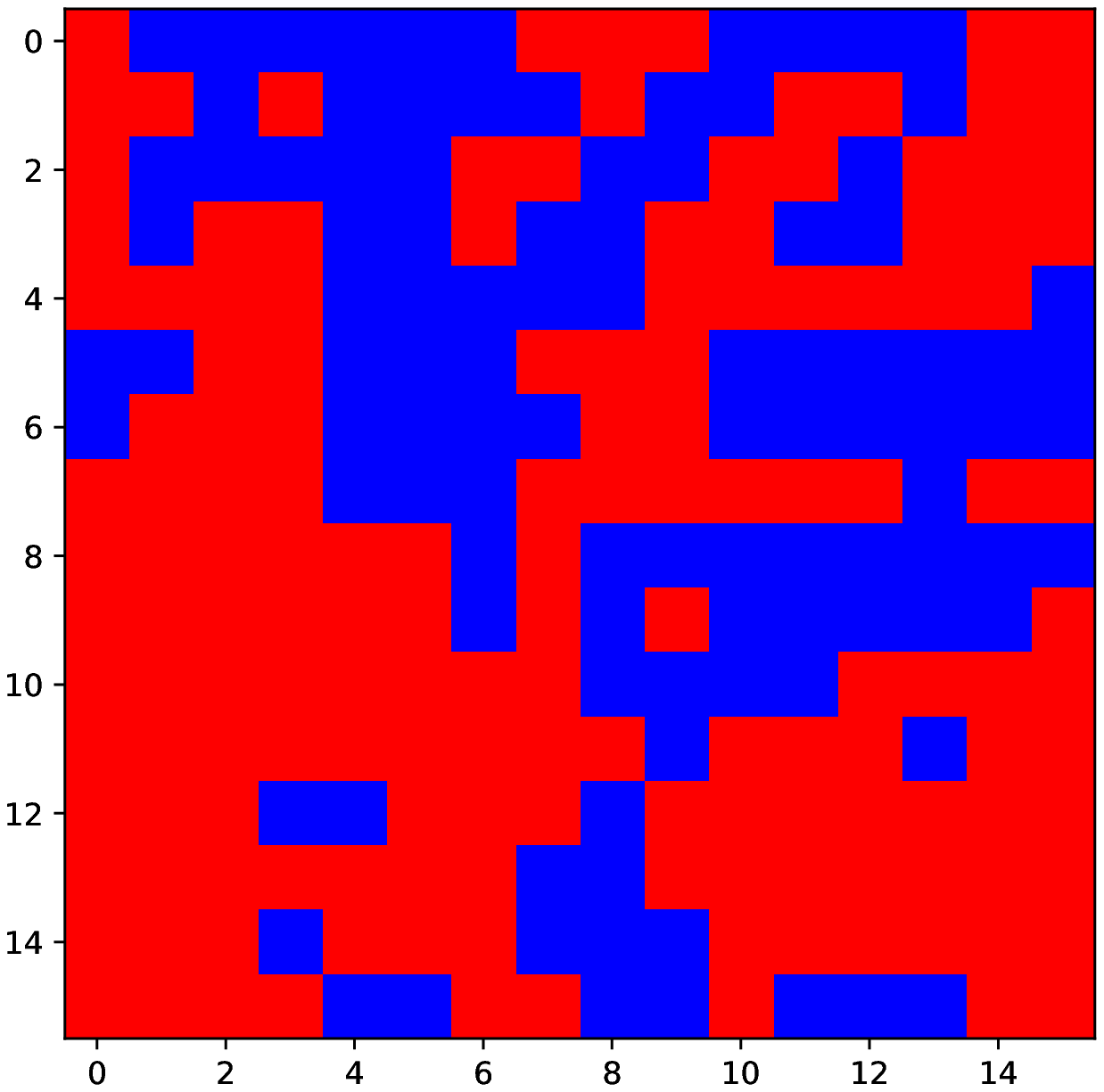}
\includegraphics[width=5.6cm]{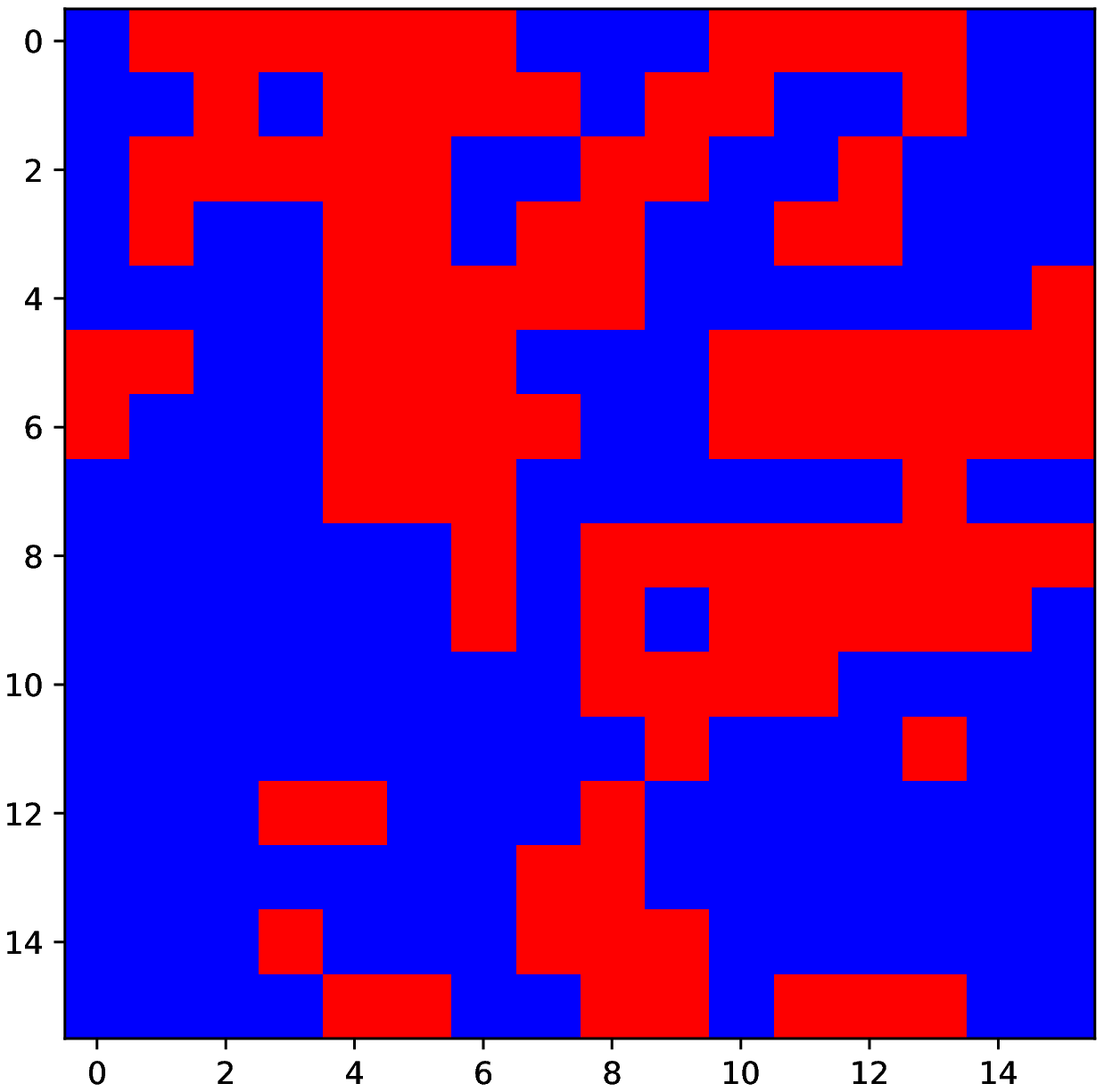}
\includegraphics[width=5.6cm]{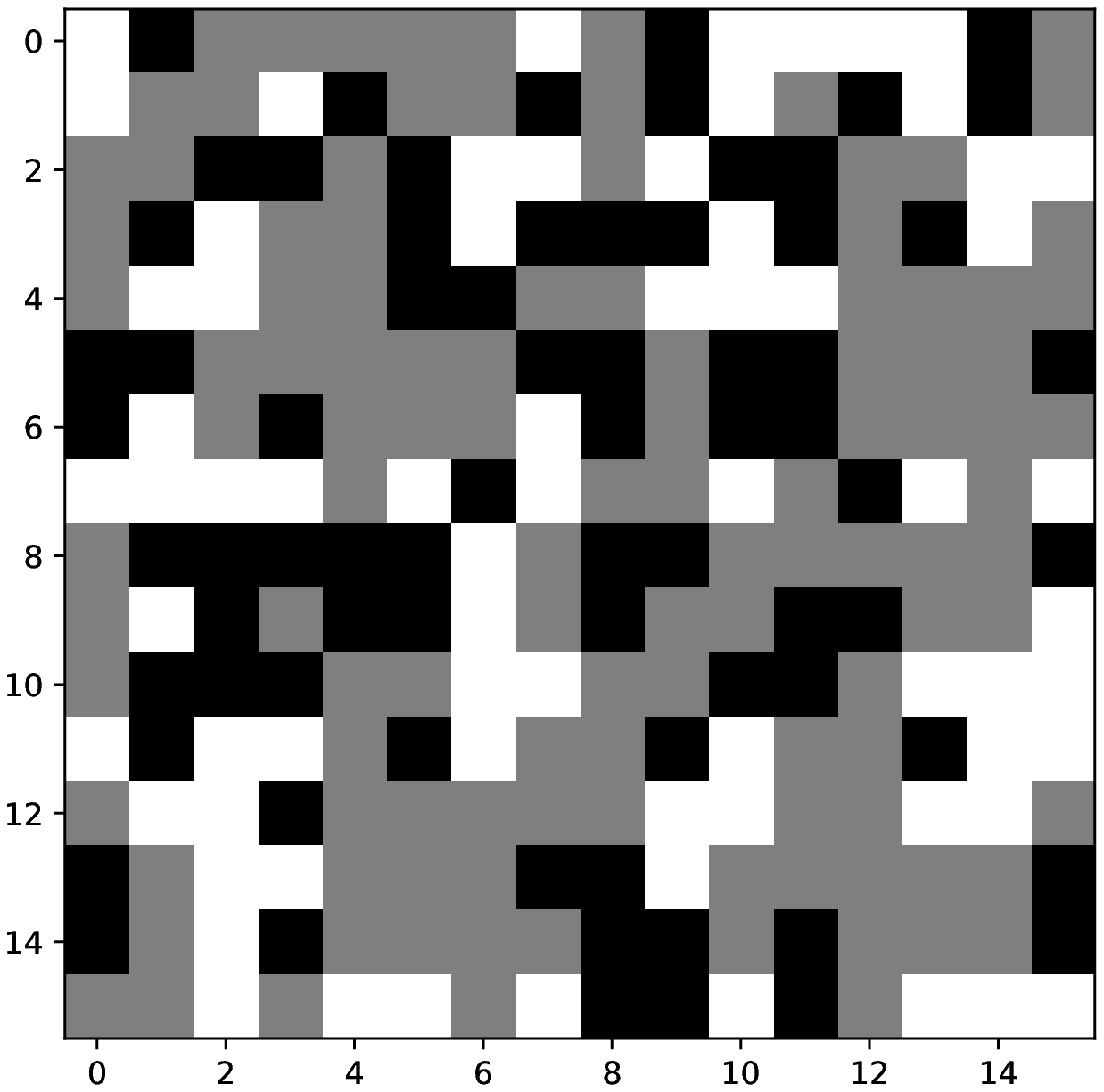}
\caption{
Examples of the spin configuration $\{ s_i \}$ (a-b, d-e) and 
correlation configuration $\{ g_i(L/2) \}$ (c, f) of the 2D Ising model. 
The upper figures (a-c) are snapshots at the low temperature of $T=2.0$, 
and the lower figures (d-f) are those at the high temperature of $T=2.8$. 
}
\label{fig:config_Ising}
\end{center}
\end{figure*}

As a supplementary information, we present examples of 
the spin configurations $\{ s_i \}$ and correlation 
configurations $\{ g_i(L/2) \}$, which is defined by 
Eq.~(7) (main text). 
The 2D Ising model is displayed 
in Fig.~S\ref{fig:config_Ising}. 
The spin configurations at the low temperature of $T=2.0$, in units 
of $J$, are given in Figs.~S\ref{fig:config_Ising}a and 
S\ref{fig:config_Ising}b. The $\pm 1$ spins are displayed 
in red and blue. These two configurations are identical 
because of the inversion symmetry. 
The corresponding correlation configuration is shown 
in Fig.~S\ref{fig:config_Ising}c. 
The correlations from $+1$ to $-1$ are mapped in gray scale 
from 255 (white) to 0 (black). 
The spin configurations at the high temperature of $T=2.8$ are 
given in Figs.~S\ref{fig:config_Ising}d-S\ref{fig:config_Ising}e, 
and the corresponding correlation configuration 
is given in Fig.~S\ref{fig:config_Ising}f. 

\begin{figure*}
\begin{center}

{\bf a} \hspace*{5.6cm}{\bf b}\hspace*{5.6cm}{\bf c}\hspace*{3.6cm}

\vspace*{1mm}

\includegraphics[width=5.6cm]{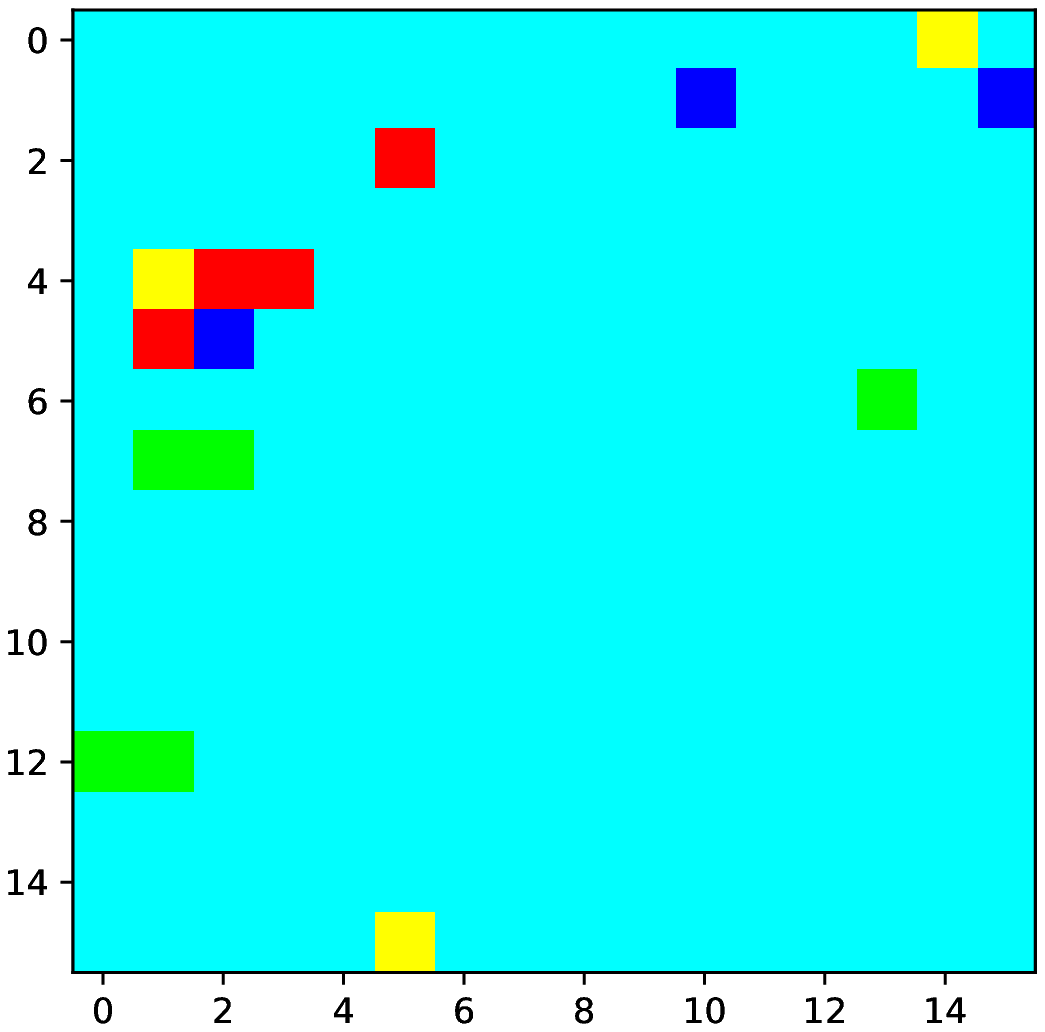}
\includegraphics[width=5.6cm]{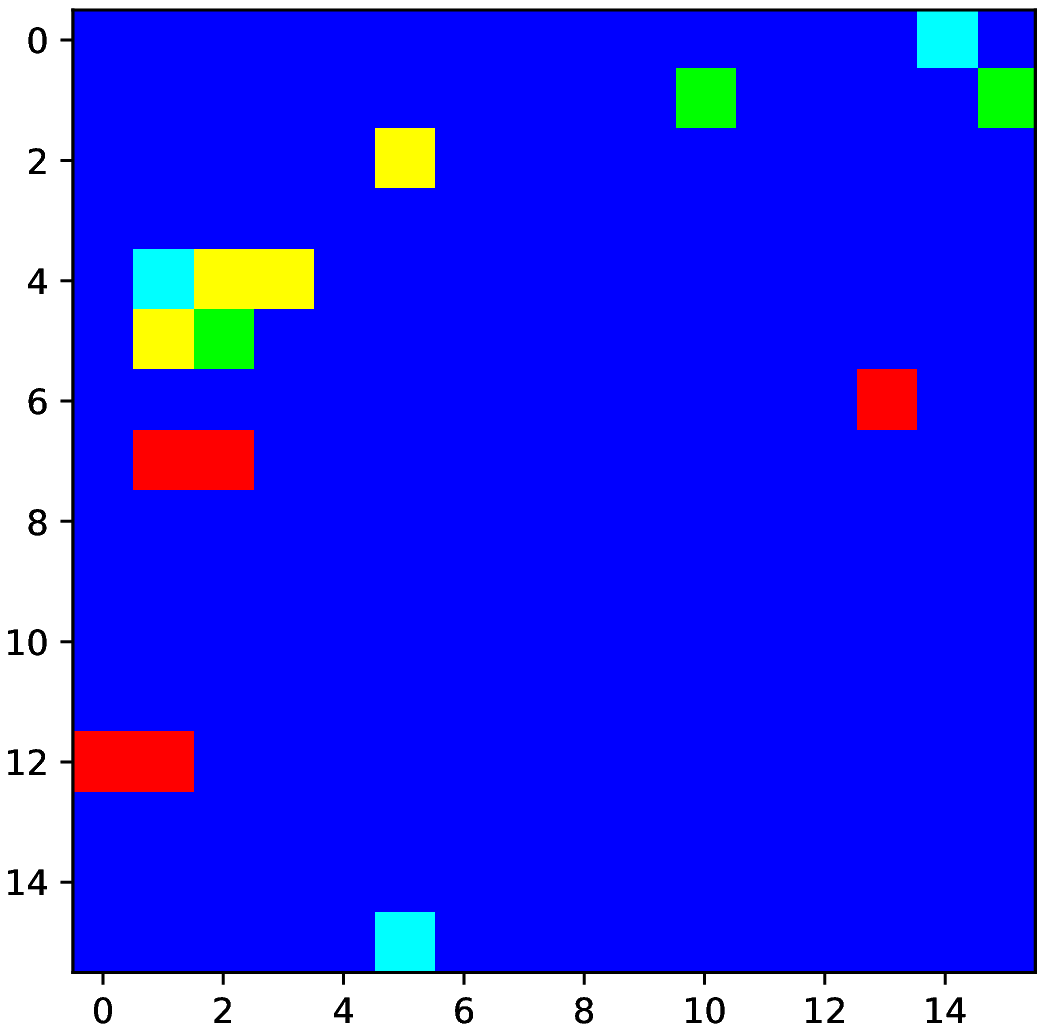}
\includegraphics[width=5.6cm]{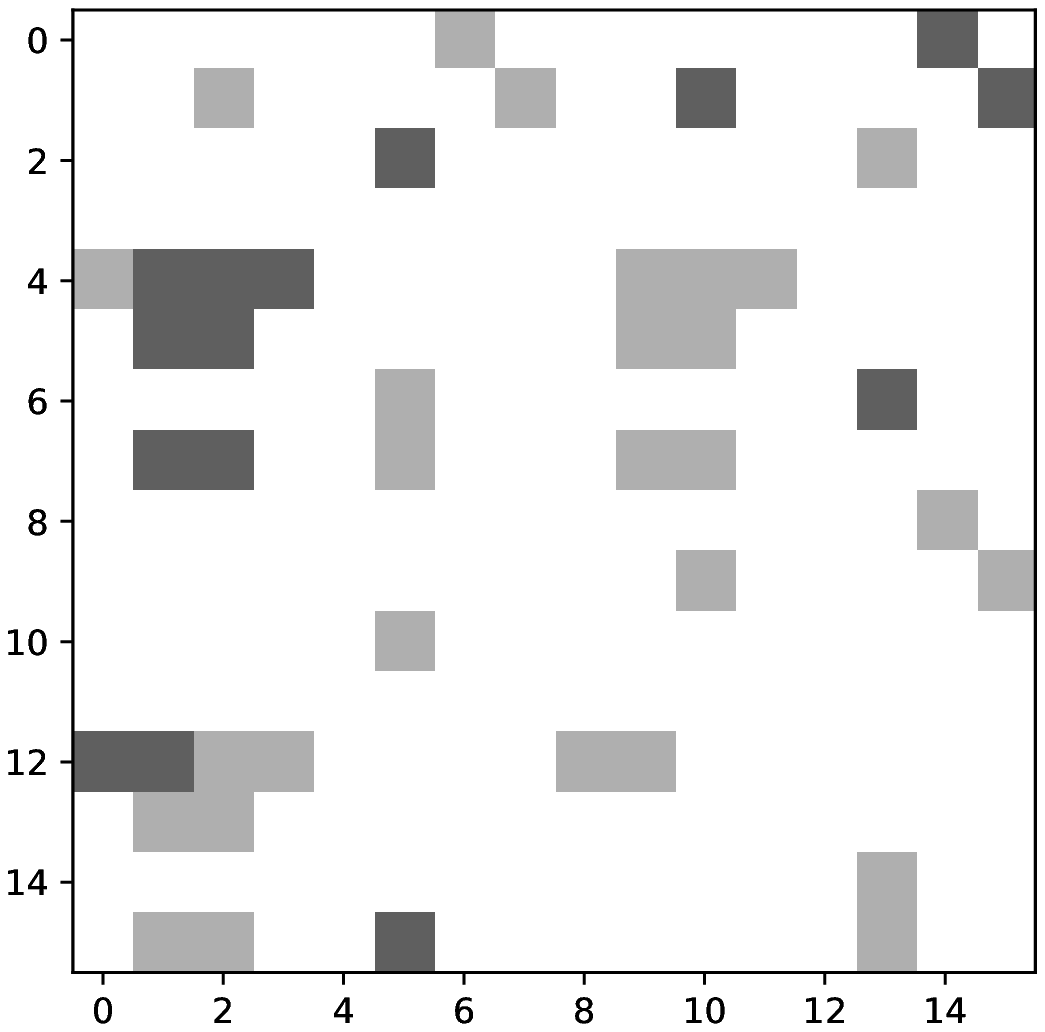}

\vspace*{2mm}

{\bf d} \hspace*{5.6cm}{\bf e}\hspace*{5.6cm}{\bf f}\hspace*{3.6cm}

\vspace*{1mm}

\includegraphics[width=5.6cm]{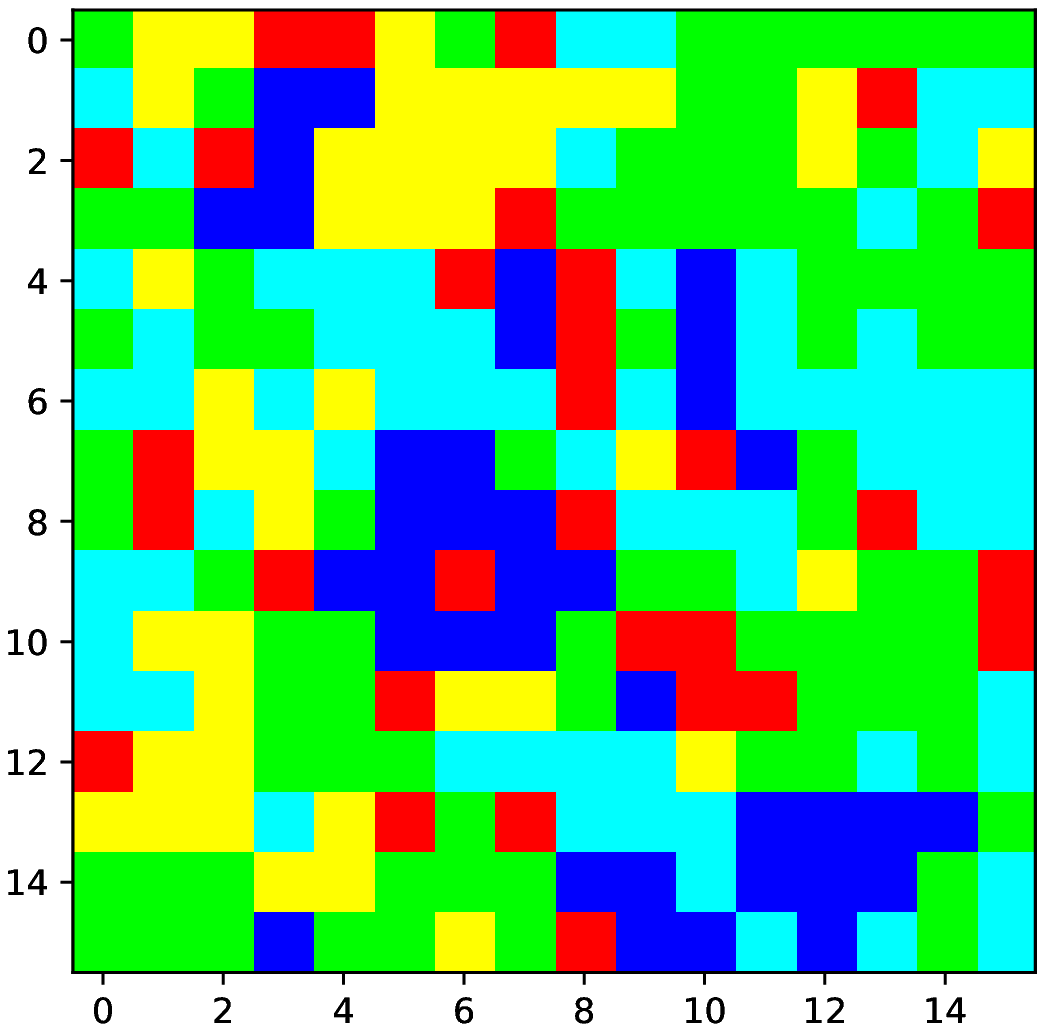}
\includegraphics[width=5.6cm]{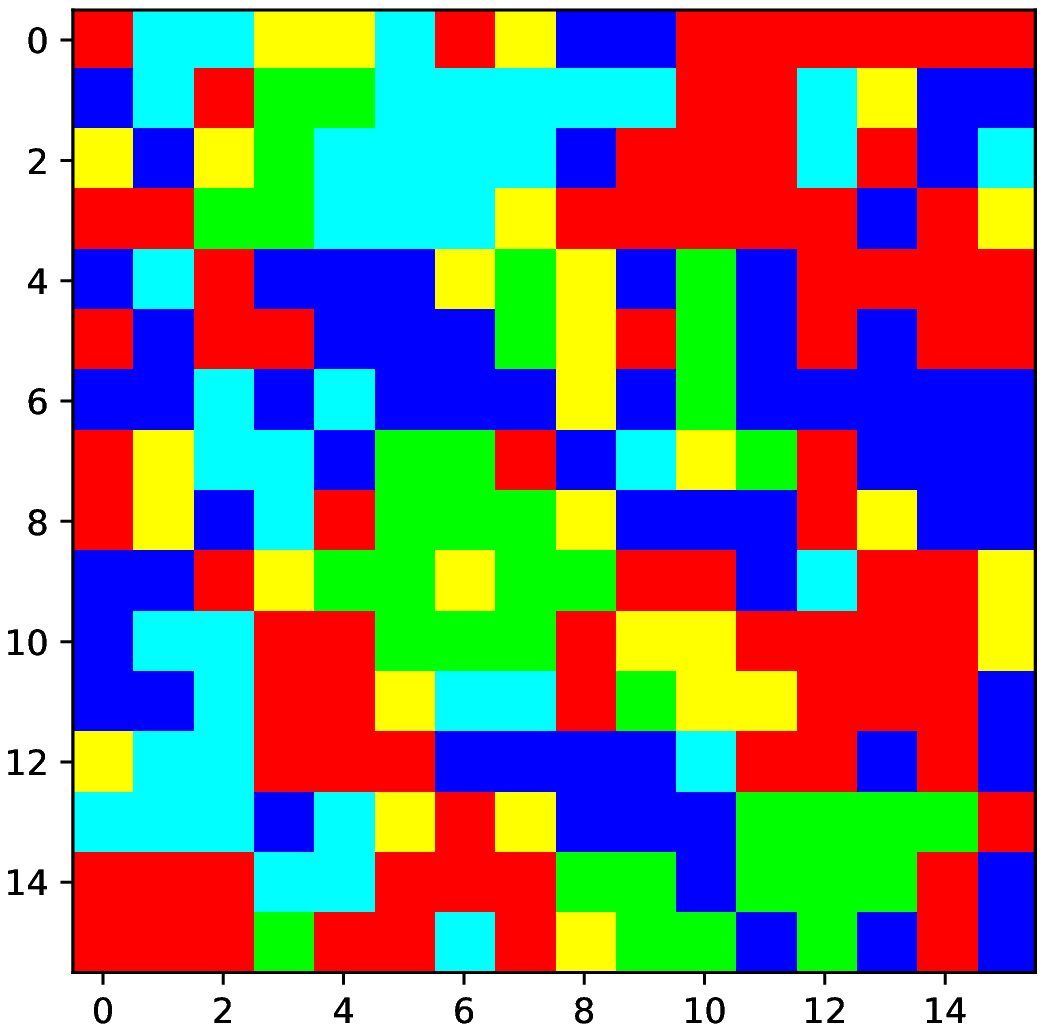}
\includegraphics[width=5.6cm]{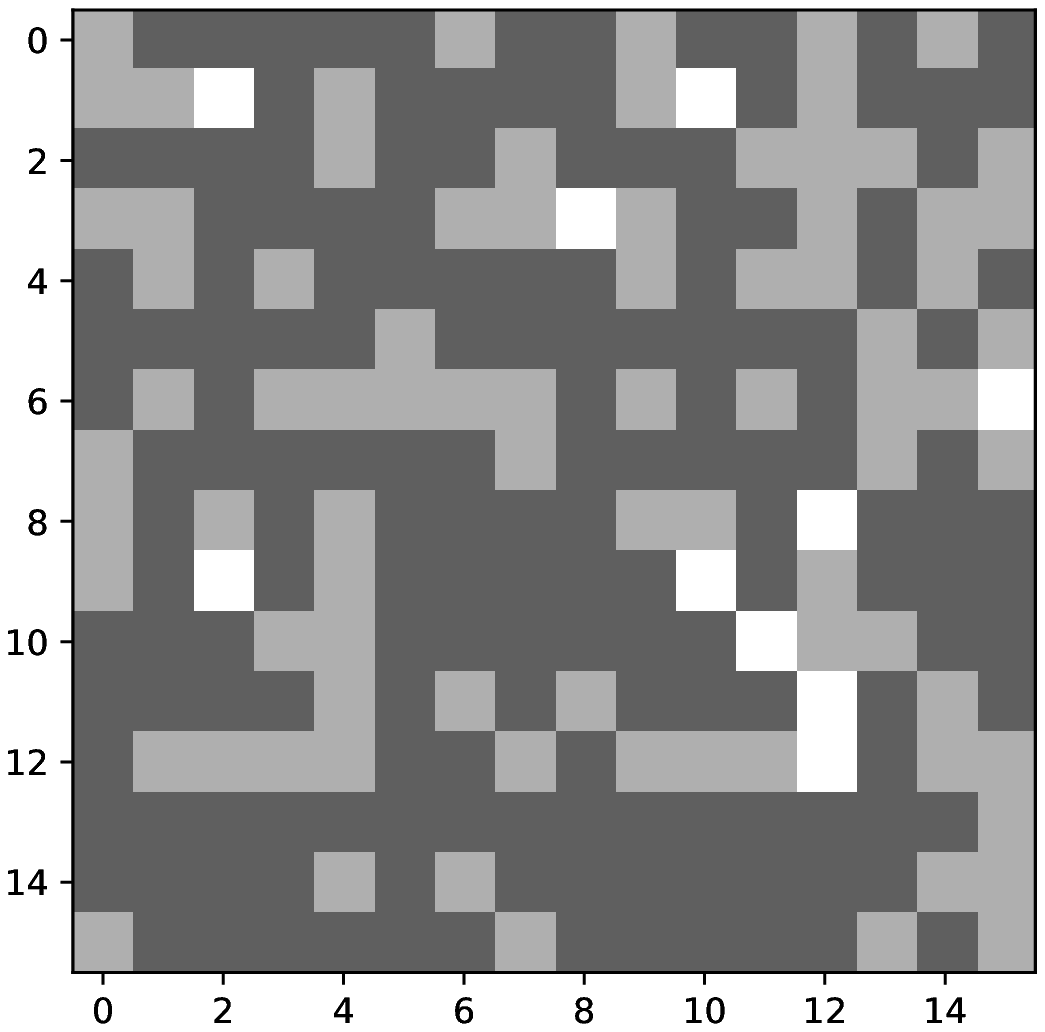}
\caption{
Examples of the spin configuration $\{ s_i \}$ (a-b, d-e) and 
correlation configuration $\{ g_i(L/2) \}$ (c, f) 
of the 2D 5-state Potts model. 
The upper figures (a-c) are snapshots at the low temperature of $T=0.8$, 
and the lower figures (d-f) are those at the high temperature of $T=1.2$. 
}
\label{fig:config_Potts}
\end{center}
\end{figure*}

Examples of the spin and correlation configurations of the 2D 
5-state Potts model are shown in Fig.~S\ref{fig:config_Potts}. 
The five spin states are displayed in five colors, and 
there is a 120(=$5!$)-fold permutational symmetry 
in the $5$-state Potts model. 
The spin configurations at the low temperature of $T=0.8$ 
are shown in Figs.~S\ref{fig:config_Potts}a-S\ref{fig:config_Potts}b, 
and those at the high temperature of $T=1.2$ 
are shown in Figs.~S\ref{fig:config_Potts}d-S\ref{fig:config_Potts}e.  
The spin configurations presented 
in Figs.~S\ref{fig:config_Potts}a and S\ref{fig:config_Potts}b 
are identical, and the corresponding correlation configuration 
is given in Fig.~S\ref{fig:config_Potts}c.  
The correlation configuration presented in Fig.~S\ref{fig:config_Potts}f 
corresponds to the spin configurations presented 
in Figs.~S\ref{fig:config_Potts}d-S\ref{fig:config_Potts}e. 
We note that the ordered-phase correlation configuration 
of the Ising model (Fig.~S\ref{fig:config_Ising}c) and 
that of the Potts model (Fig.~S\ref{fig:config_Potts}c) 
are similar. At the same time, the behavior of the 
disordered-phase correlation configuration 
of the Ising model (Fig.~S\ref{fig:config_Ising}f) and 
that of the Potts model (Fig.~S\ref{fig:config_Potts}f) 
are similar. 

\begin{figure*}
\begin{center}

{\bf a} \hspace*{5.6cm}{\bf b}\hspace*{5.6cm}{\bf c}\hspace*{3.6cm}

\vspace*{1mm}

\includegraphics[width=5.6cm]{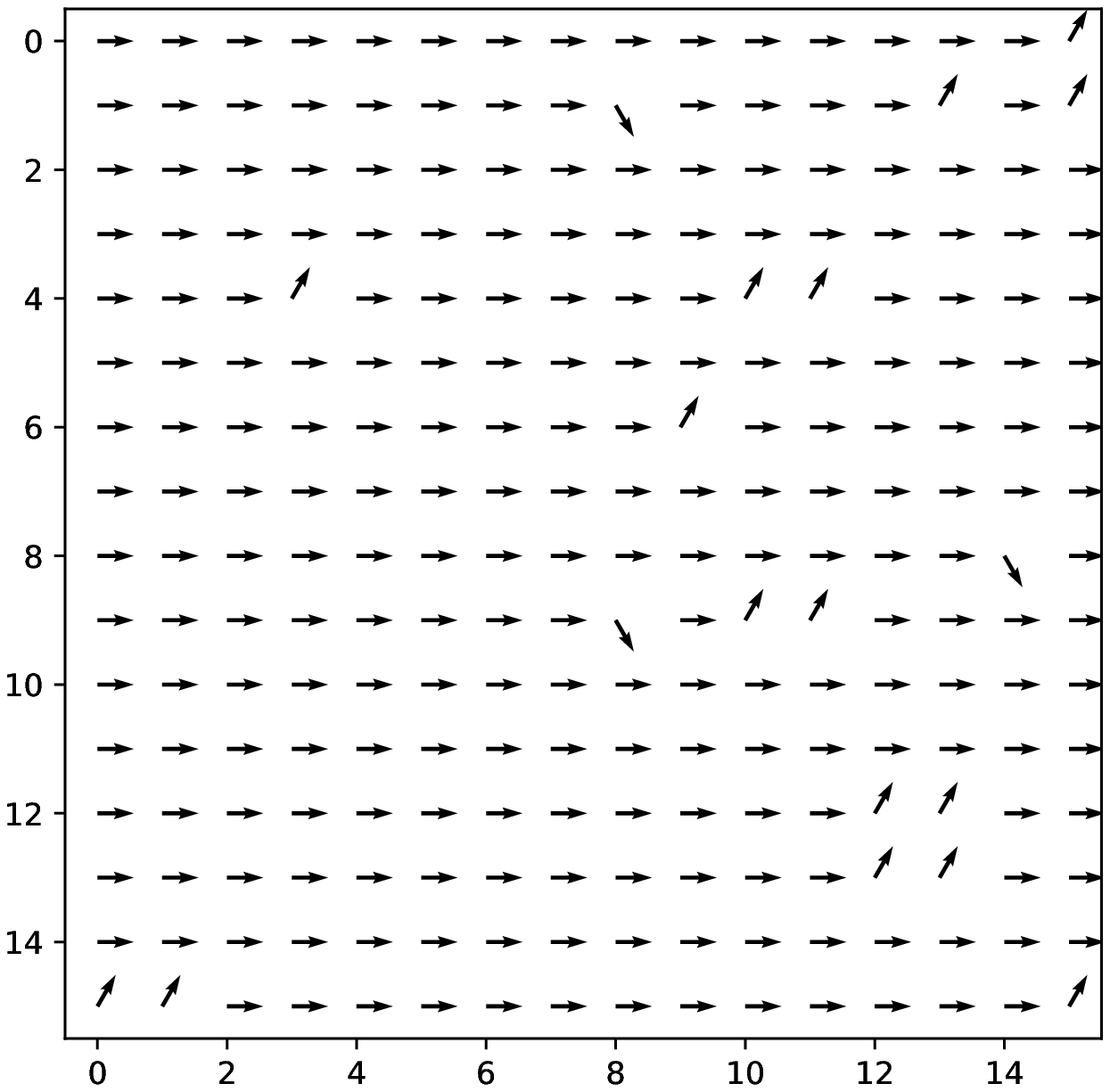}
\includegraphics[width=5.6cm]{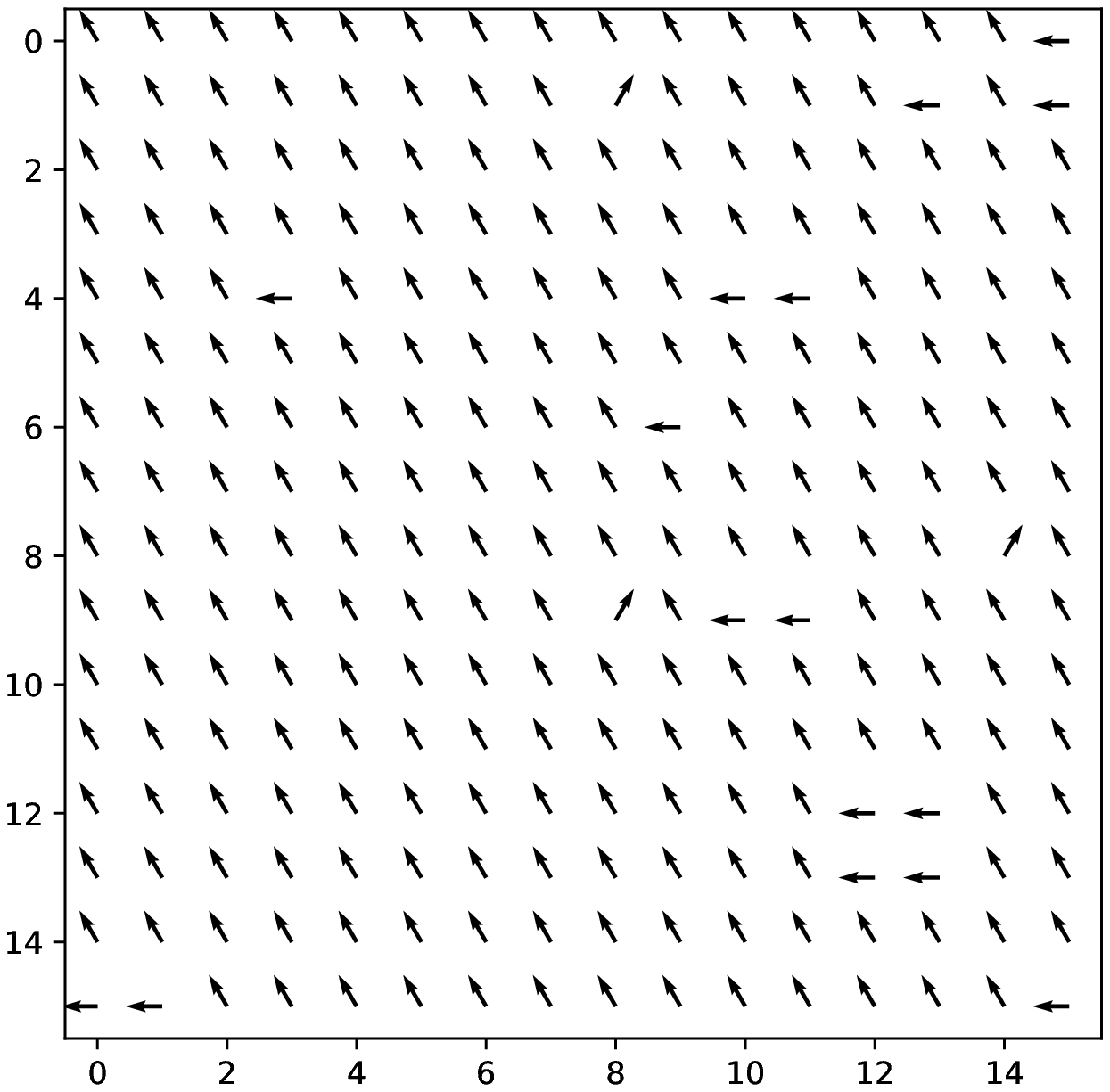}
\includegraphics[width=5.6cm]{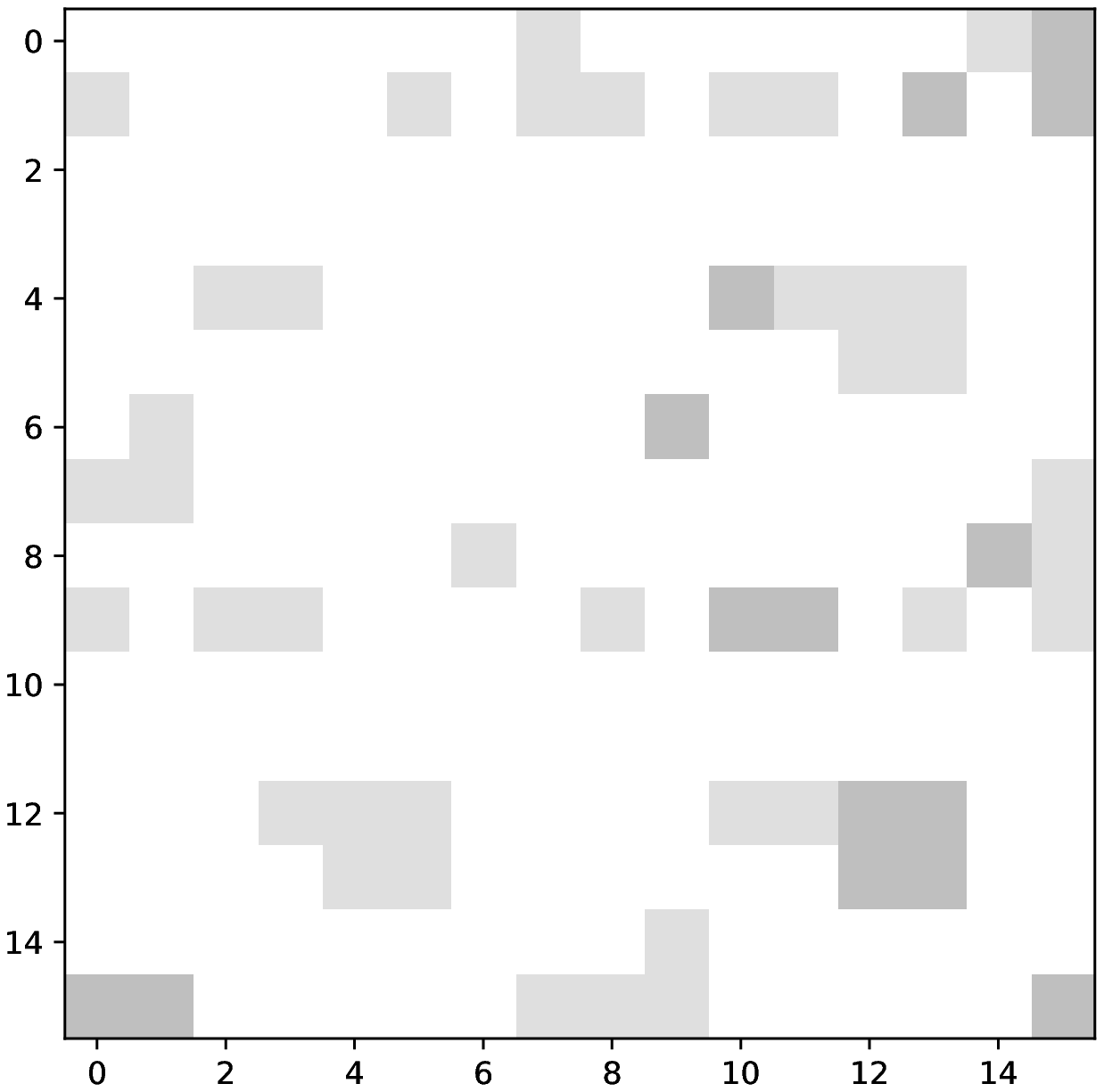}

\vspace*{2mm}

{\bf d} \hspace*{5.6cm}{\bf e}\hspace*{5.6cm}{\bf f}\hspace*{3.6cm}

\vspace*{1mm}

\includegraphics[width=5.6cm]{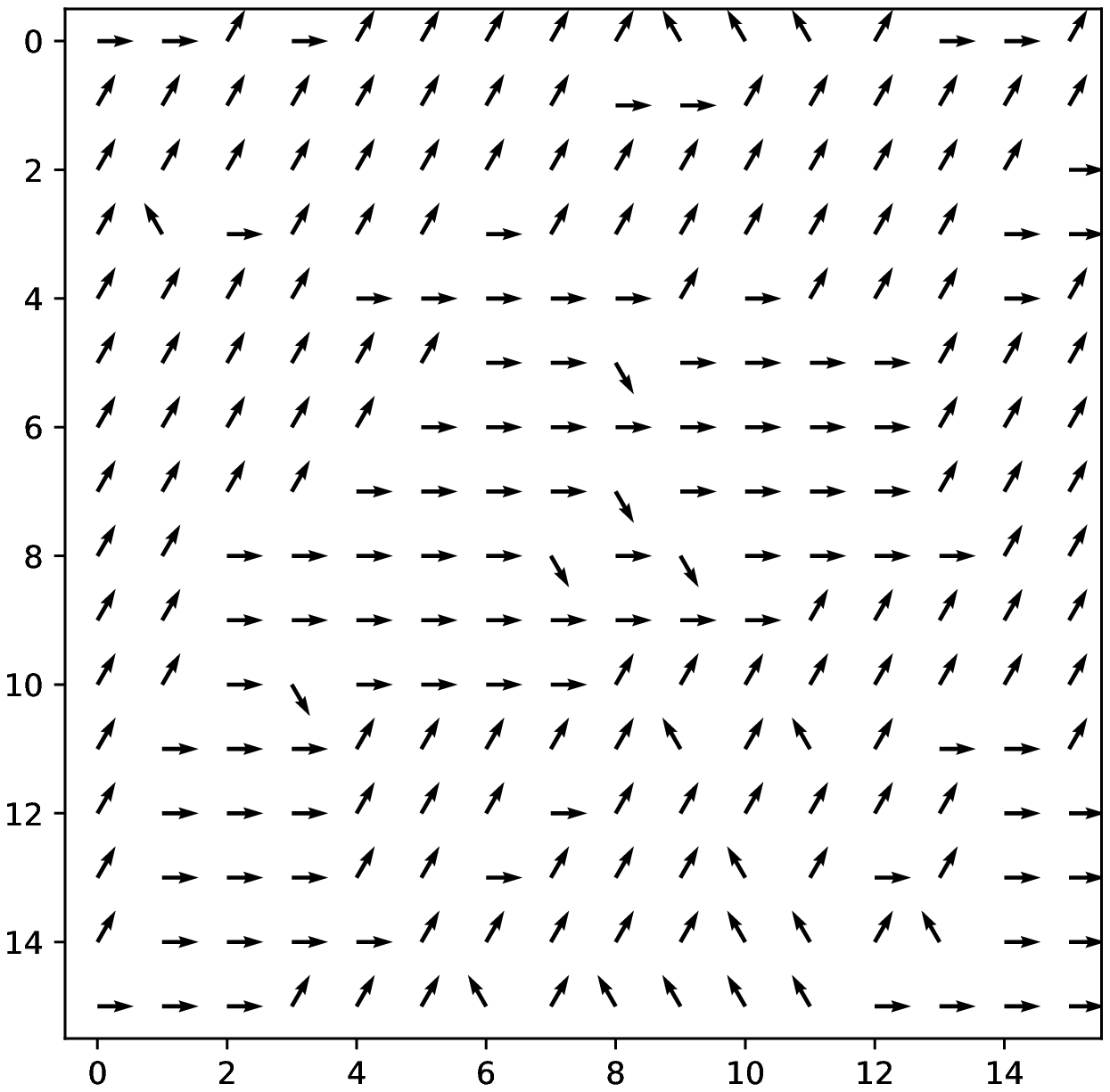}
\includegraphics[width=5.6cm]{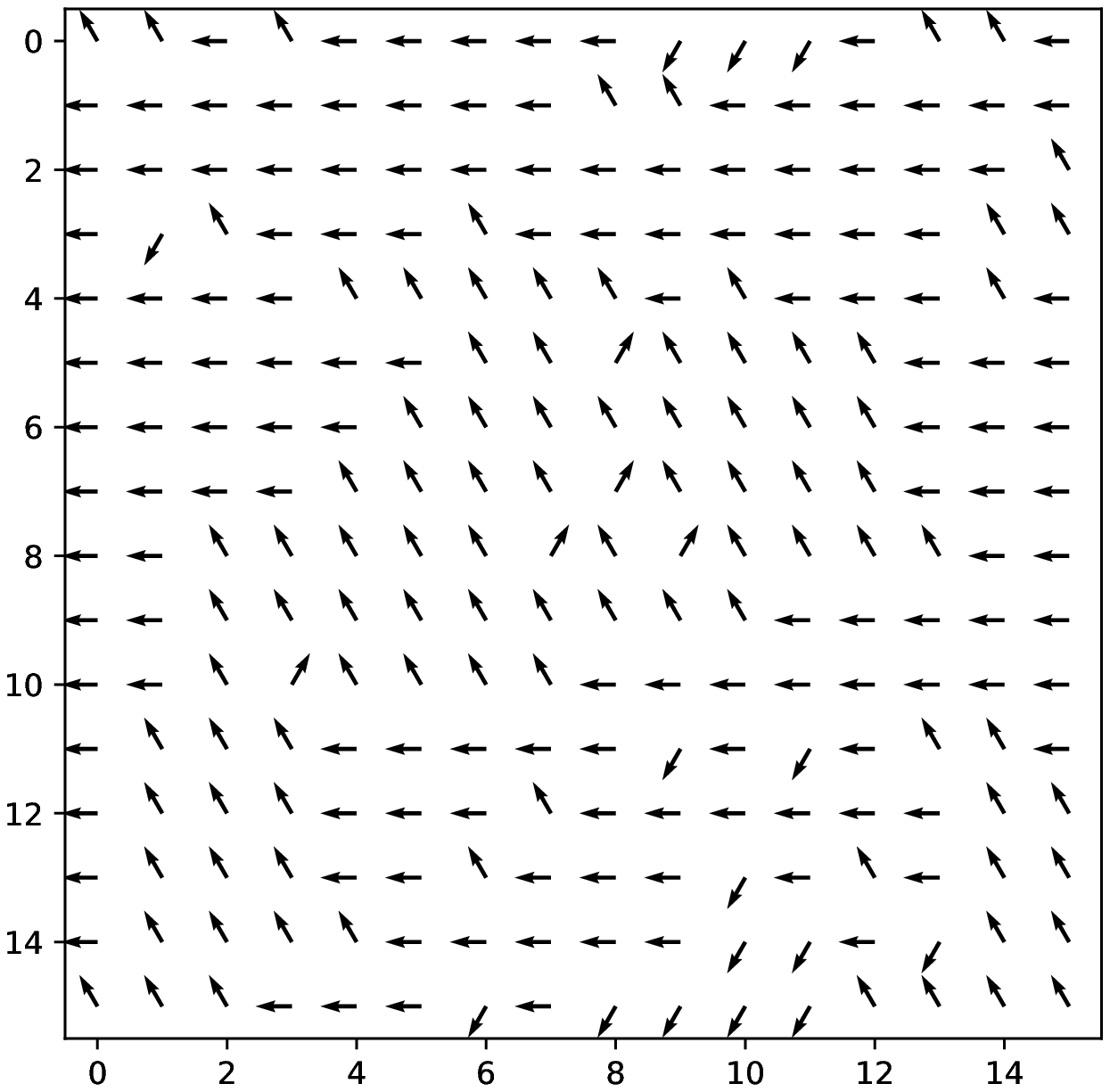}
\includegraphics[width=5.6cm]{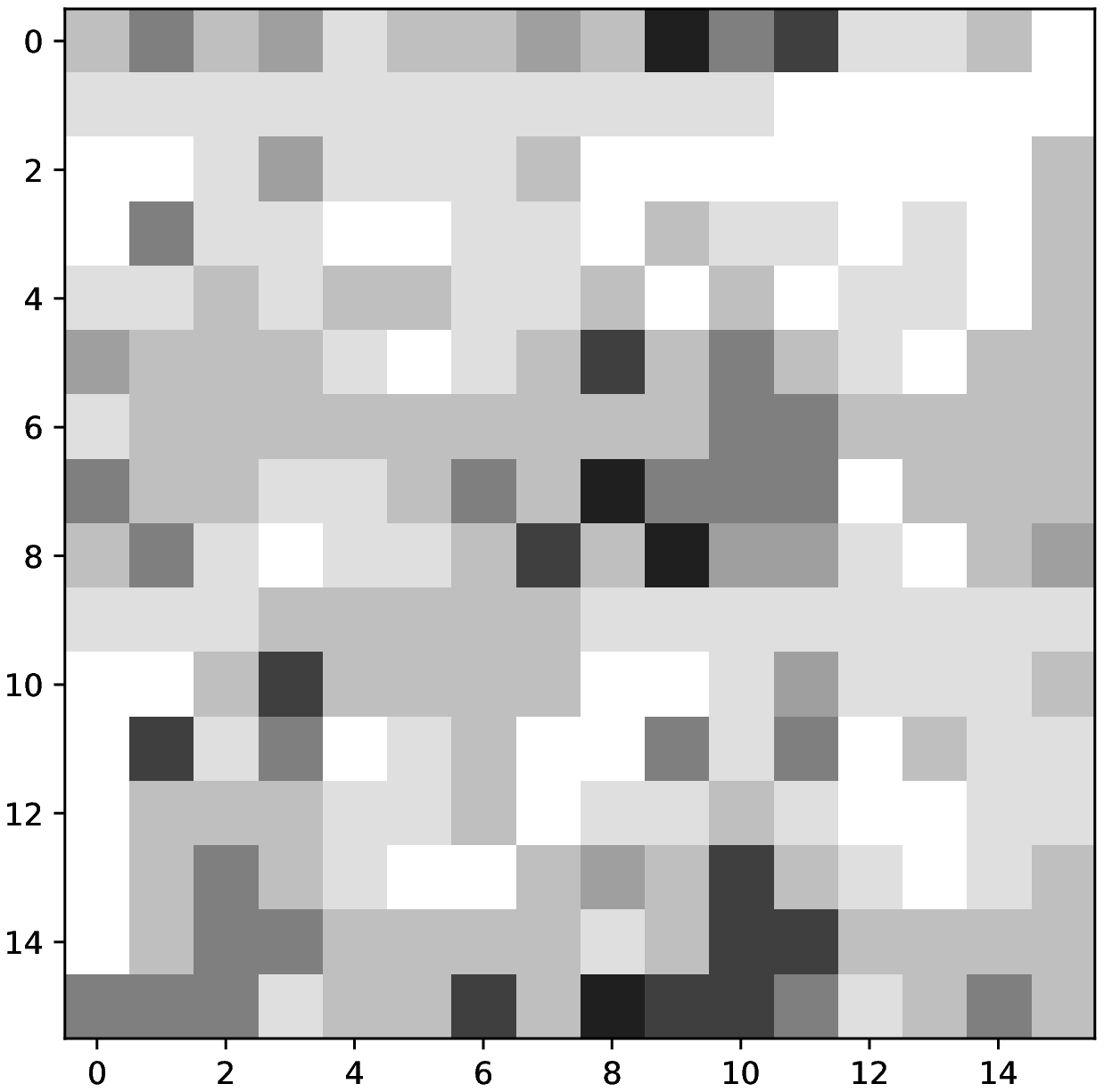}

\vspace*{2mm}

{\bf g} \hspace*{5.6cm}{\bf h}\hspace*{5.6cm}{\bf i}\hspace*{3.6cm}

\vspace*{1mm}

\includegraphics[width=5.6cm]{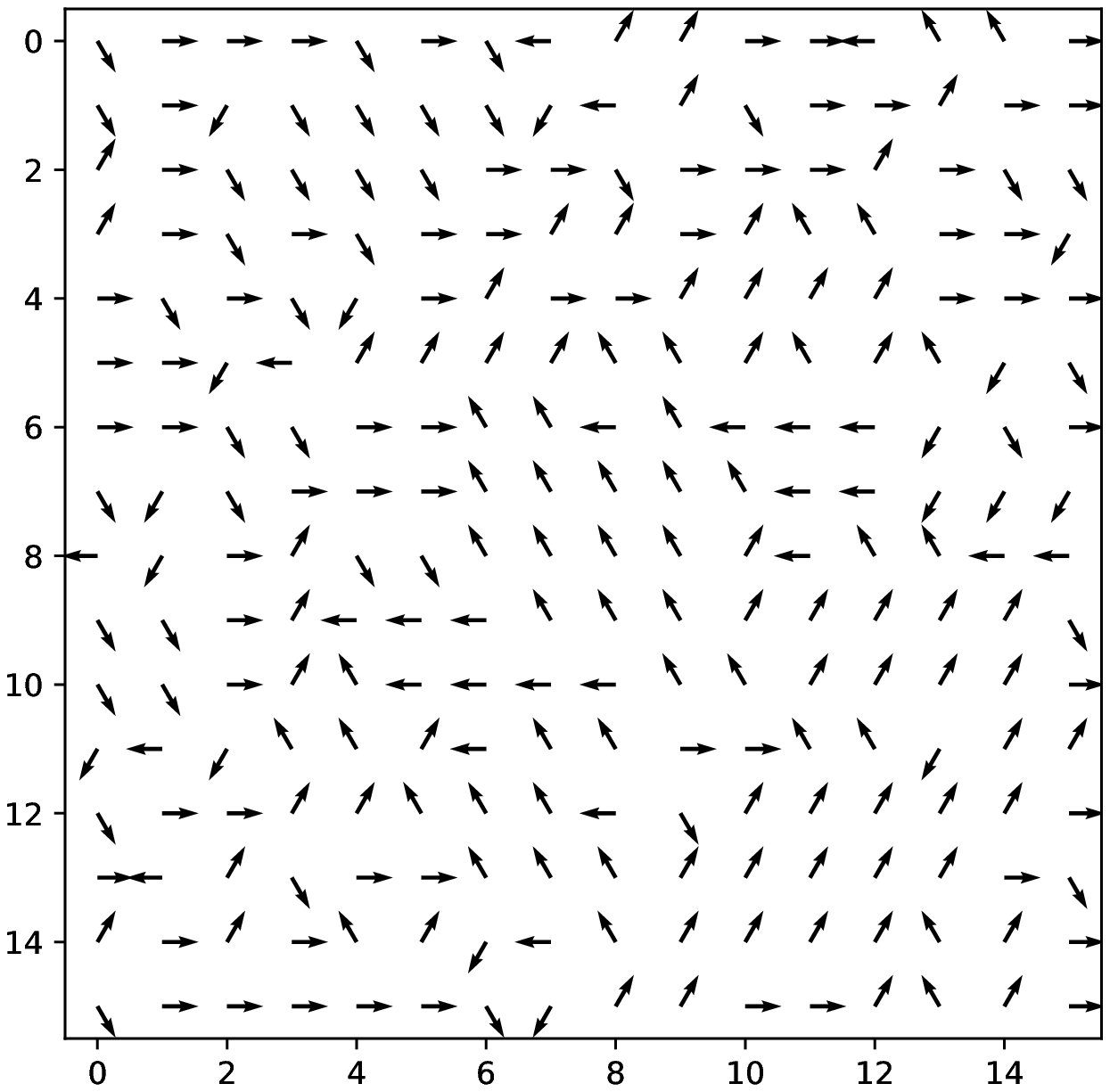}
\includegraphics[width=5.6cm]{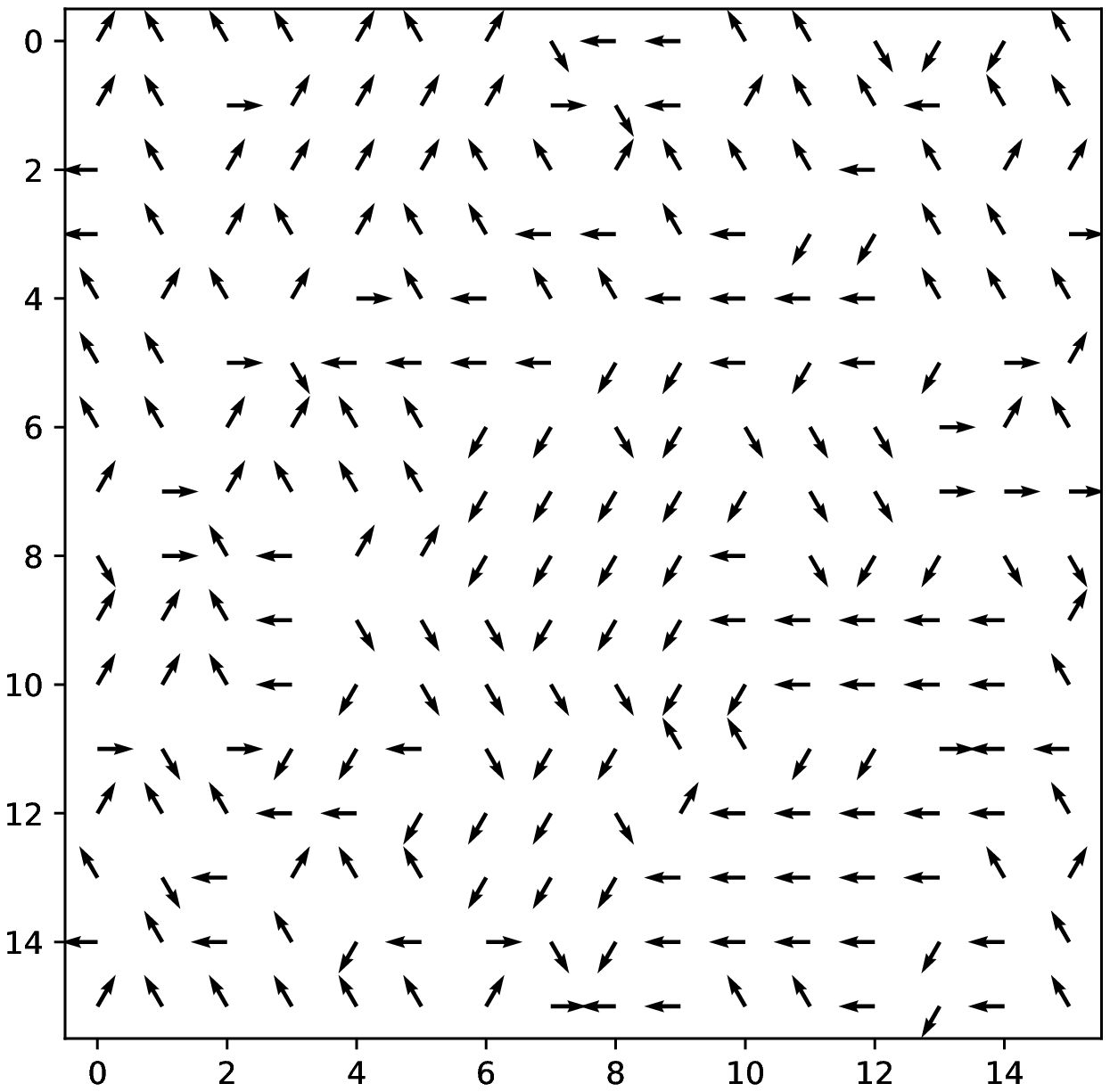}
\includegraphics[width=5.6cm]{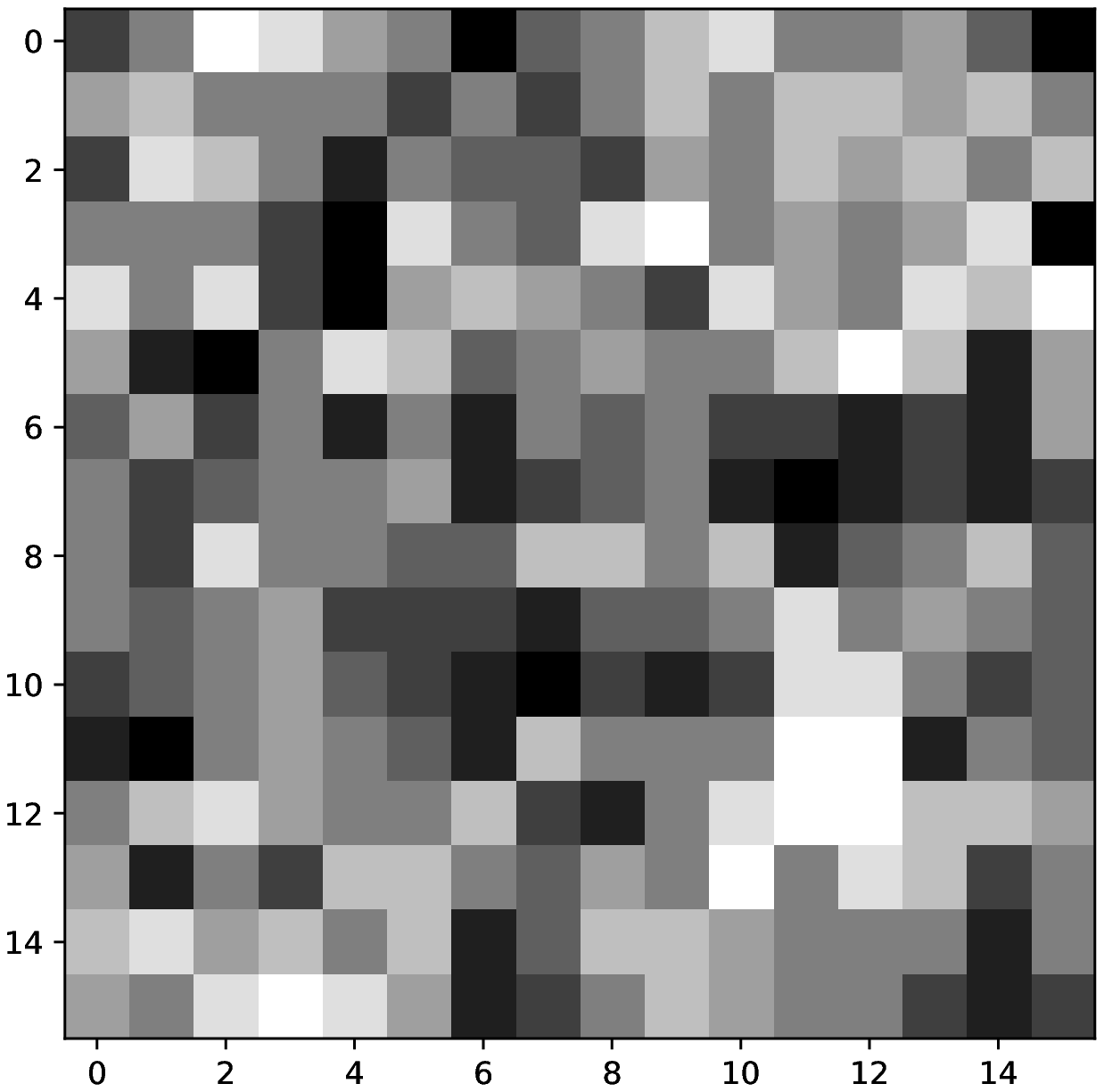}
\caption{
Examples of the spin configuration $\{ \theta_i \}$ (a-b, d-e, g-h) and 
correlation configuration $\{ g_i(L/2) \}$ (c, f, i) 
of the 2D 6-state clock model. 
The upper figures (a-c) are snapshots at the low temperature of $T=0.5$, 
the middle figures (d-f) are those at the mid-range temperature of $T=0.8$, 
and the lower figures (g-i) are those at the high temperature of $T=1.2$. 
}
\label{fig:config_clock}
\end{center}
\end{figure*}

We have presented the spin and correlation configurations of 
the 2D 6-state clock model in Fig.~S\ref{fig:config_clock}.
There is a 6-fold rotational symmetry. 
The spin configurations at the low temperature of $T=0.5$, 
at the mid-range temperature of $T=0.8$, and at the high temperature 
of $T=1.2$ are given 
in Figs.~S\ref{fig:config_clock}a-S\ref{fig:config_clock}b, 
S\ref{fig:config_clock}d-S\ref{fig:config_clock}e, and 
S\ref{fig:config_clock}g-S\ref{fig:config_clock}h, respectively. 
The corresponding correlation configurations are given 
in Figs.~S\ref{fig:config_clock}c, S\ref{fig:config_clock}f, 
and S\ref{fig:config_clock}i, respectively. 

For all the models, the correlation configurations, shown in 
the third column of figures, exhibit similar behavior;  
they are almost white at low temperatures, whereas at high temperatures, 
they are mixtures of white, gray, and black.

